\documentclass[a4paper,11pt]{article}
\pdfoutput=1
\usepackage{jcappub}

\usepackage{amsmath}
\usepackage{bm}
\usepackage{xcolor,color,colortbl}
\usepackage[utf8]{inputenc}
\usepackage{hyperref,orcidlink}
\usepackage{url}
\usepackage{graphicx}
\usepackage{multirow,bigdelim}
\usepackage{amssymb}
\usepackage[amssymb]{SIunits}
\usepackage{multirow}
\usepackage{bbold}
\usepackage{verbatim}

\definecolor{darkgreen}{HTML}{339933}

\usepackage[normalem]{ulem}

\newcommand{\mfon}{m_{419}} 
\newcommand{\mfoo}{m_{501}} 
\newcommand{\msst}{m_{673}} 

\title{The clustering of Lyman Alpha Emitting galaxies at $z=2-3$}

\author[1]{{Martin~White}\orcidlink{0000-0001-9912-5070},}
\author[2]{{A.~Raichoor}\orcidlink{0000-0001-5999-7923},}
\author[3]{{Arjun~Dey}\orcidlink{0000-0002-4928-4003},}
\author[4,5]{{Lehman~H.~Garrison}\orcidlink{0000-0002-9853-5673},}
\author[6]{{Eric~Gawiser}\orcidlink{0000-0003-1530-8713},}
\author[7]{{D.~Lang},}
\author[8]{{Kyoung-soo~Lee}\orcidlink{0000-0003-3004-9596},}
\author[9]{{A.~D.~Myers},}
\author[2]{{D.~Schlegel},}
\author[3]{{F.~Valdes}\orcidlink{0000-0001-5567-1301},}
\author[2]{{J.~Aguilar},}
\author[10]{{S.~Ahlen}\orcidlink{0000-0001-6098-7247},}
\author[11]{{D.~Brooks},}
\author[2]{{E.~Chaussidon}\orcidlink{0000-0001-8996-4874},}
\author[2]{{T.~Claybaugh},}
\author[12]{{K.~Dawson},}
\author[13]{{A.~de la Macorra}\orcidlink{0000-0002-1769-1640},}
\author[14]{{Biprateep~Dey}\orcidlink{0000-0002-5665-7912},}
\author[11]{{P.~Doel},}
\author[15,16]{{K.~Fanning}\orcidlink{0000-0003-2371-3356},}
\author[11,17]{{A.~Font-Ribera}\orcidlink{0000-0002-3033-7312},}
\author[18,19]{{J.~E.~Forero-Romero}\orcidlink{0000-0002-2890-3725},}
\author[2]{{S.~Gontcho A Gontcho}\orcidlink{0000-0003-3142-233X},}
\author[20]{{G.~Gutierrez},}
\author[2]{{J.~Guy}\orcidlink{0000-0001-9822-6793},}
\author[21,22,23]{{K.~Honscheid},}
\author[24]{{D.~Kirkby}\orcidlink{0000-0002-8828-5463},}
\author[2]{{A.~Kremin}\orcidlink{0000-0001-6356-7424},}
\author[2]{{M.~Landriau}\orcidlink{0000-0003-1838-8528},}
\author[25]{{L.~Le~Guillou}\orcidlink{0000-0001-7178-8868},}
\author[2]{{M.~E.~Levi}\orcidlink{0000-0003-1887-1018},}
\author[26]{{C.~Magneville},}
\author[27,17]{{M.~Manera}\orcidlink{0000-0003-4962-8934},}
\author[21,28,23]{{P.~Martini}\orcidlink{0000-0002-4279-4182},}
\author[3]{{A.~Meisner}\orcidlink{0000-0002-1125-7384},}
\author[29,17]{{R.~Miquel},}
\author[30]{{B.~Moon}\orcidlink{0009-0008-4022-3870},}
\author[14]{{J.~ A.~Newman}\orcidlink{0000-0001-8684-2222},}
\author[31,32]{{G.~Niz}\orcidlink{0000-0002-1544-8946},}
\author[26,2]{{N.~Palanque-Delabrouille}\orcidlink{0000-0003-3188-784X},}
\author[30]{{C.~Park}\orcidlink{0000-0003-3078-2763},}
\author[33,7,34]{{W.~J.~Percival}\orcidlink{0000-0002-0644-5727},}
\author[35]{{F.~Prada}\orcidlink{0000-0001-7145-8674},}
\author[36]{{G.~Rossi},}
\author[26]{{V.~Ruhlmann-Kleider}\orcidlink{0009-0000-6063-6121},}
\author[37]{{E.~Sanchez}\orcidlink{0000-0002-9646-8198},}
\author[38]{{E.~F.~Schlafly}\orcidlink{0000-0002-3569-7421},}
\author[39,40]{{M.~Schubnell},}
\author[41]{{H.~Seo}\orcidlink{0000-0002-6588-3508},}
\author[3]{{D.~Sprayberry},}
\author[40]{{G.~Tarl\'{e}}\orcidlink{0000-0003-1704-0781},}
\author[3]{{B.~A.~Weaver},}
\author[30]{{Y.~Yang}\orcidlink{0000-0003-3078-2763},}
\author[26]{{C.~Yèche}\orcidlink{0000-0001-5146-8533},}
\author[42]{{H.~Zou}\orcidlink{0000-0002-6684-3997},}

\affiliation[1]{Department of Physics, University of California, Berkeley, 366 LeConte Hall MC 7300, Berkeley, CA 94720-7300, USA}
\affiliation[2]{Lawrence Berkeley National Laboratory, 1 Cyclotron Road, Berkeley, CA 94720, USA}
\affiliation[3]{NSF NOIRLab, 950 N. Cherry Ave., Tucson, AZ 85719, USA}
\affiliation[4]{Center for Computational Astrophysics, Flatiron Institute, 162 5\textsuperscript{th} Avenue, New York, NY 10010, USA}
\affiliation[5]{Scientific Computing Core, Flatiron Institute, 162 5\textsuperscript{th} Avenue, New York, NY 10010, USA}
\affiliation[6]{Department of Physics and Astronomy, Rutgers, The State University of New Jersey, Piscataway, NJ 08854, USA}
\affiliation[7]{Perimeter Institute for Theoretical Physics, 31 Caroline St. North, Waterloo, ON N2L 2Y5, Canada}
\affiliation[8]{Department of Physics and Astronomy, Purdue University, 525 Northwestern Ave., West Lafayette, IN 47906, USA}
\affiliation[9]{Department of Physics \& Astronomy, University  of Wyoming, 1000 E. University, Dept.~3905, Laramie, WY 82071, USA}
\affiliation[10]{Physics Dept., Boston University, 590 Commonwealth Avenue, Boston, MA 02215, USA}
\affiliation[11]{Department of Physics \& Astronomy, University College London, Gower Street, London, WC1E 6BT, UK}
\affiliation[12]{Department of Physics and Astronomy, The University of Utah, 115 South 1400 East, Salt Lake City, UT 84112, USA}
\affiliation[13]{Instituto de F\'{\i}sica, Universidad Nacional Aut\'{o}noma de M\'{e}xico,  Cd. de M\'{e}xico  C.P. 04510,  M\'{e}xico}
\affiliation[14]{Department of Physics \& Astronomy and Pittsburgh Particle Physics, Astrophysics, and Cosmology Center (PITT PACC), University of Pittsburgh, 3941 O'Hara Street, Pittsburgh, PA 15260, USA}
\affiliation[15]{Kavli Institute for Particle Astrophysics and Cosmology, Stanford University, Menlo Park, CA 94305, USA}
\affiliation[16]{SLAC National Accelerator Laboratory, Menlo Park, CA 94305, USA}
\affiliation[17]{Institut de F\'{i}sica d’Altes Energies (IFAE), The Barcelona Institute of Science and Technology, Campus UAB, 08193 Bellaterra Barcelona, Spain}
\affiliation[18]{Departamento de F\'isica, Universidad de los Andes, Cra. 1 No. 18A-10, Edificio Ip, CP 111711, Bogot\'a, Colombia}
\affiliation[19]{Observatorio Astron\'omico, Universidad de los Andes, Cra. 1 No. 18A-10, Edificio H, CP 111711 Bogot\'a, Colombia}
\affiliation[20]{Fermi National Accelerator Laboratory, PO Box 500, Batavia, IL 60510, USA}
\affiliation[21]{Center for Cosmology and AstroParticle Physics, The Ohio State University, 191 West Woodruff Avenue, Columbus, OH 43210, USA}
\affiliation[22]{Department of Physics, The Ohio State University, 191 West Woodruff Avenue, Columbus, OH 43210, USA}
\affiliation[23]{The Ohio State University, Columbus, 43210 OH, USA}
\affiliation[24]{Department of Physics and Astronomy, University of California, Irvine, 92697, USA}
\affiliation[25]{Sorbonne Universit\'{e}, CNRS/IN2P3, Laboratoire de Physique Nucl\'{e}aire et de Hautes Energies (LPNHE), FR-75005 Paris, France}
\affiliation[26]{IRFU, CEA, Universit\'{e} Paris-Saclay, F-91191 Gif-sur-Yvette, France}
\affiliation[27]{Departament de F\'{i}sica, Serra H\'{u}nter, Universitat Aut\`{o}noma de Barcelona, 08193 Bellaterra (Barcelona), Spain}
\affiliation[28]{Department of Astronomy, The Ohio State University, 4055 McPherson Laboratory, 140 W 18th Avenue, Columbus, OH 43210, USA}
\affiliation[29]{Instituci\'{o} Catalana de Recerca i Estudis Avan\c{c}ats, Passeig de Llu\'{\i}s Companys, 23, 08010 Barcelona, Spain}
\affiliation[30]{Korea Astronomy and Space Science Institute, 776 Daedeokdae-ro, Yuseong-gu, Daejeon 34055, Republic of Korea}
\affiliation[31]{Departamento de F\'{i}sica, Universidad de Guanajuato - DCI, C.P. 37150, Leon, Guanajuato, M\'{e}xico}
\affiliation[32]{Instituto Avanzado de Cosmolog\'{\i}a A.~C., San Marcos 11 - Atenas 202. Magdalena Contreras, 10720. Ciudad de M\'{e}xico, M\'{e}xico}
\affiliation[33]{Department of Physics and Astronomy, University of Waterloo, 200 University Ave W, Waterloo, ON N2L 3G1, Canada}
\affiliation[34]{Waterloo Centre for Astrophysics, University of Waterloo, 200 University Ave W, Waterloo, ON N2L 3G1, Canada}
\affiliation[35]{Instituto de Astrof\'{i}sica de Andaluc\'{i}a (CSIC), Glorieta de la Astronom\'{i}a, s/n, E-18008 Granada, Spain}
\affiliation[36]{Department of Physics and Astronomy, Sejong University, Seoul, 143-747, Korea}
\affiliation[37]{CIEMAT, Avenida Complutense 40, E-28040 Madrid, Spain}
\affiliation[38]{Space Telescope Science Institute, 3700 San Martin Drive, Baltimore, MD 21218, USA}
\affiliation[39]{Department of Physics, University of Michigan, Ann Arbor, MI 48109, USA}
\affiliation[40]{University of Michigan, Ann Arbor, MI 48109, USA}
\affiliation[41]{Department of Physics \& Astronomy, Ohio University, Athens, OH 45701, USA}
\affiliation[42]{National Astronomical Observatories, Chinese Academy of Sciences, A20 Datun Rd., Chaoyang District, Beijing, 100012, P.R. China}

\abstract{We measure the clustering of Lyman Alpha Emitting galaxies (LAEs) selected from the One-hundred-square-degree DECam Imaging in Narrowbands (ODIN) survey, with spectroscopic follow-up from Dark Energy Spectroscopic Instrument (DESI).  We use DESI spectroscopy to optimize our selection and to constrain the interloper fraction and redshift distribution of our narrow-band selected sources.  We select samples of 4000 LAEs at $z=2.45$ and $3.1$ in 9 sq.deg.\ centered on the COSMOS field with median Ly$\alpha$ fluxes of $\approx 10^{-16}\,\mathrm{erg}\,\mathrm{s}^{-1}\mathrm{cm}^{-2}$.  Covariances and cosmological inferences are obtained from a series of mock catalogs built upon high-resolution N-body simulations that match the footprint, number density, redshift distribution and observed clustering of the sample.  We find that both samples have a correlation length of $r_0=3.0\pm 0.2\,h^{-1}$Mpc.  Within our fiducial cosmology these correspond to 3D number densities of $\approx 10^{-3}\,h^3\,\mathrm{Mpc}^{-3}$ and, from our mock catalogs, biases of 1.7 and 2.0 at $z=2.45$ and $3.1$, respectively.  We discuss the implications of these measurements for the use of LAEs as large-scale structure tracers for high-redshift cosmology.}

\begin{document}
\maketitle
\flushbottom

\section{Introduction}
\label{sec:introduction}

The inhomogeneous Universe, as probed by fluctuations in the cosmic microwave background (CMB) radiation or surveys of the large-scale structure of the Universe, provides one of our best windows on fundamental physics at ultra-high energies \cite{SnowmassCF}.  The tightest constraints on dark energy, mass limits on light dark matter particles, models of inflation, neutrino masses and light relic particles all come from one or both of these measurements.  There are compelling theoretical motivations \cite{Wilson19,Sailer21,Ferraro22} to push the study of large-scale structure to redshifts $2<z<6$ using both relativistic and non-relativistic tracers.  This will allow us to probe the metric, particle content and \emph{both} epochs of accelerated expansion (Inflation and Dark Energy domination) with high precision in a regime that is not theory limited.

A high redshift survey that aims to measure the large-scale structure in three dimensions needs to be able to efficiently obtain redshifts for faint galaxies over wide areas.  While we are witnessing tremendous advances in instrumental capability, this is still challenging for faint galaxies and at high redshift unless the galaxies that are targeted have bright emission lines.  One such population of galaxies are Lyman Alpha Emitters (LAEs; \cite{Ouchi20}), which -- as their names suggests -- have prominent Ly$\alpha$ emission lines in their spectra.  Existing surveys are somewhat limited -- cosmic variance is a major concern and many questions remain unsettled -- but studies suggest that LAEs have relatively low stellar masses, low star formation rates, young ages and low dust content \cite{Ouchi20}.  If LAEs populate low halo masses (compared to other galaxies that might be selected for spectroscopic follow-up at that redshift) then we might expect that they have a low and roughly scale-independent bias.  Under such conditions, LAEs would make an excellent visible tracer of the underlying matter distribution.

Motivated by the possibility of using large-scale structure at early times as a cosmological probe \cite{Wilson19,Sailer21,Ferraro22}, the Dark Energy Spectroscopic Instrument (DESI; \cite{Levi13,DESI}) collaboration took spectra of LAE candidate galaxies selected from various surveys as part of ancillary\footnote{For DESI collaboration studies of the broader class of Lyman Break Galaxies see ref.~\cite{Kleider24}.} programs. In this paper, we present our measurements of the clustering of LAEs selected from the One-hundred-deg$^2$ DECam Imaging in Narrowbands (ODIN; \cite{Lee24,Ramakrishnan23,Firestone24}) survey.  
A subset of these objects were followed up spectroscopically by DESI, with the resulting data used to optimize the narrow-band selection and constrain the interloper/outlier fraction and the redshift distribution, $dN/dz$.  After a brief introduction to the data employed (\S\ref{sec:data}; described further in our companion paper, ref.~\cite{odin_sample_selection}) we describe the clustering analysis (\S\ref{sec:clustering}).  Our pipeline is tested, covariance matrices computed and inferences are obtained using a series of mock catalogs built upon high resolution N-body simulations.  These are described in \S\ref{sec:mocks}.  Our main results are given in \S\ref{sec:results} and the implications for surveys aiming to use LAEs for high-$z$ cosmology are presented in \S\ref{sec:forecasts}.  Finally we conclude in \S\ref{sec:conclusions}, with some technical details relegated to an appendix.

\section{Data}
\label{sec:data}

In this section we give a brief overview of the datasets underlying our analysis.  These data are drawn from two surveys: the One-hundred-square-degree DECam Imaging in Narrowbands (ODIN; \cite{Lee24,Ramakrishnan23,Firestone24}) and the Dark Energy Spectroscopic Instrument (DESI; \cite{Levi13,DESI}).  We briefly describe these data below, referring the reader to the preceeding references and companion papers for more details.

\subsection{The ODIN survey}

ODIN (NOIRLab Survey Program 2020B-0201) is a wide-field ($\sim 100\,\mathrm{deg}^2$), deep, imaging survey targeting seven fields in three narrow-band filters tuned to select LAEs at redshifts of $z\approx 2.4$, 3.1, and 4.5. Covering a (comoving) volume of $\approx 0.24\,\mathrm{Gpc}^3$, comparable to the Sloan Digital Sky Survey \cite{SDSSI}, the ODIN sample is designed to measure the large-scale clustering of LAEs across cosmological time. The ODIN survey is described in detail in ref.~\cite{Lee24}; here, we just summarize the relevant details.

The survey narrow-band observations were obtained between the 2021A-2023B semesters using the Dark Energy Camera \citep[DECam;][]{DECam} on the Victor M. Blanco 4-meter telescope at the Cerro Tololo Inter-American Observatory in Chile. The narrow-band imaging depths of 25.5, 25.7, and 25.9 AB mag ($5\,\sigma$ in a $2^{\prime\prime}$ diameter aperture) in the N419, N501 and N673 filters, respectively, correspond to Ly$\alpha$ line flux limits of $3.1\times10^{-17}$, $1.8\times10^{-17}$, $1.1\times10^{-17}\, {\rm erg\, s^{-1}\,cm^{-2}}$ at redshifts of $z\simeq 2.45, 3.1$ and 4.5 respectively. The narrow-band ODIN observations are complemented with existing public broad-band imaging data from the DESI Legacy Imaging Survey (LS; \cite{DECaLS}) and the Subaru Hyper-Suprime Cam (HSC) Strategic Survey Program (SSP; \cite{HSCSSP}). 

Approximately half of the ODIN fields overlap the survey footprint of the DESI spectroscopic survey, and two (COSMOS and XMM) have been targeted by DESI for spectroscopic observations of candidate LAEs. The selection and DESI spectrosopic observations of the candidate LAEs are complex and are discussed fully in a companion paper \cite{odin_sample_selection}, but here we provide a brief summary. At the time the DESI spectroscopic observations were initiated, the ODIN observations were still in progress. Due to scheduling and resource constraints, we were only able to properly spectroscopically characterize LAE candidates in the COSMOS field (N419 and N501 candidates) and the XMM field (N419 candidates). In this paper, we focus only on the N419 and N501 LAE candidates in the COSMOS field. 

\begin{table}[t]
\centering
\begin{tabular}{lcccccc}
\hline
 Filter & Area & $N_{\rm targ}$ & $N_{\rm obs}$ & $N_{\rm VI,ok}$ & $N_{\rm VI,zok}$ & Interloper fraction\\
 & [deg$^2$] & & & & & \\
\hline
{\bf N419} & 8.90 & 2382 & 897 & 838 & 822 & 0.02 -- 0.08\\
{\bf N501} & 9.34 & 1956 & 1240 & 1145 & 1099 & 0.04 -- 0.11\\
\hline
\end{tabular}
\caption{Selected LAE candidates in COSMOS for our `refined' sample (see text). We report for each band the number of selected candidates ($N_{\rm targ}$), the number of observed candidates ($N_{\rm obs}$), the number of those with a VI-validated redshift ($N_{\rm VI,ok}$) and with a VI-validated redshift in the relevant redshift range ($N_{\rm VI,zok}$ and $z_{\rm min} < z_{\rm VI,ok} < z_{\rm max}$). We lastly report our estimated interloper fraction range (see text for details).}
\label{tab:data}
\end{table}

We constructed photometric catalogs of the imaging data using Tractor \cite{Tractor2016}, with the narrow-band imaging data as the detection image, performing forced photometry on the broad-band data. LAE candidate sources were selected for spectroscopic targeting according to a variety of photometric cuts which were designed to be ``liberal'', in the sense of allowing a larger fraction of interloper galaxies in order to determine how best to (re)optimize the sample selection (post redshift determination) for high purity and completeness. We refer the interested reader to ref.~\cite{odin_sample_selection} for the (quite involved) details of the candidate selection for the spectroscopic observations. 

\subsection{DESI spectroscopy}

The DESI spectroscopic observations were obtained as part of an ancillary program to target candidate high-redshift galaxy populations with the goal of investigating their suitability for high-redshift clustering studies. DESI \cite{Levi13,DESI}, a prime-focus fiber-fed multi-object spectrograph mounted on the Nicholas U. Mayall 4-meter Telescope of the Kitt Peak National Observatory, has the ability to obtain simultaneous spectra of $\approx 5000$ targets within a 3~deg diameter field \cite{DESIb,DESI_Instrument_Overview,Silber23,Miller24}. DESI covers a wavelength range of $\lambda\lambda$3600-9800\AA\ with a resolution ($R\equiv\lambda/\Delta\lambda$) varying from 2200 to 5000, and the high system throughput and high efficiency make it an ideal instrument for spectroscopic redshift surveys of faint emission line sources over large contiguous fields. 

The ODIN candidates in COSMOS were observed with DESI in two campaigns \cite{odin_sample_selection}. During the first campaign, in March 2022, N501 targets were observed on a dedicated tile (82636) for an effective time\footnote{DESI uses the concept of ``effective time'' to quantify the measured squared signal-to-noise ratio in the spectra -- we refer the reader to refs.~\cite{odin_sample_selection,DESI_Spectro_Pipeline} for further details.} of 2.75 hrs. In the second campaign, in April 2023, which was part of a large pilot program, N419 and N501 sources were observed for either 1.1 hrs or 2.2 hrs. In total, we targeted approximately 3500 N419 and 3000 N501 sources in the COSMOS field, from which we selected our `refined' targets.

The spectroscopic observations were reduced by the DESI pipeline and the spectrum of each LAE candidate source was visually inspected by a team of volunteers in order to measure a redshift and assign a quality flag. These ``visual inspection'' (VI) redshifts were combined in order to create the final redshift lists for the targeted sources \cite{odin_sample_selection}.

\subsection{LAE target selections}

\begin{figure}
\centering
\begin{tabular}{cc}
    \includegraphics[width=0.45\columnwidth]{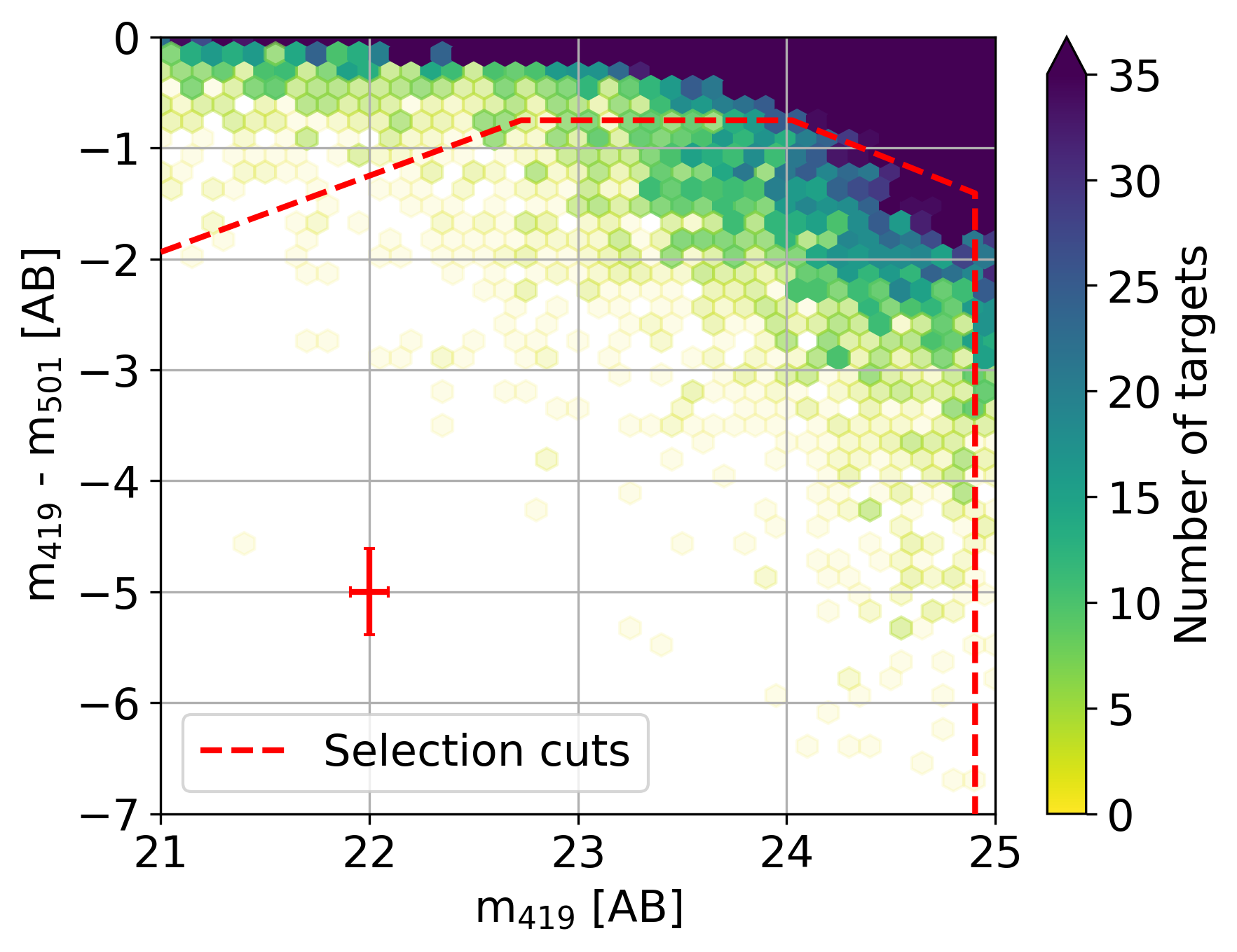} &
    \includegraphics[width=0.45\columnwidth]{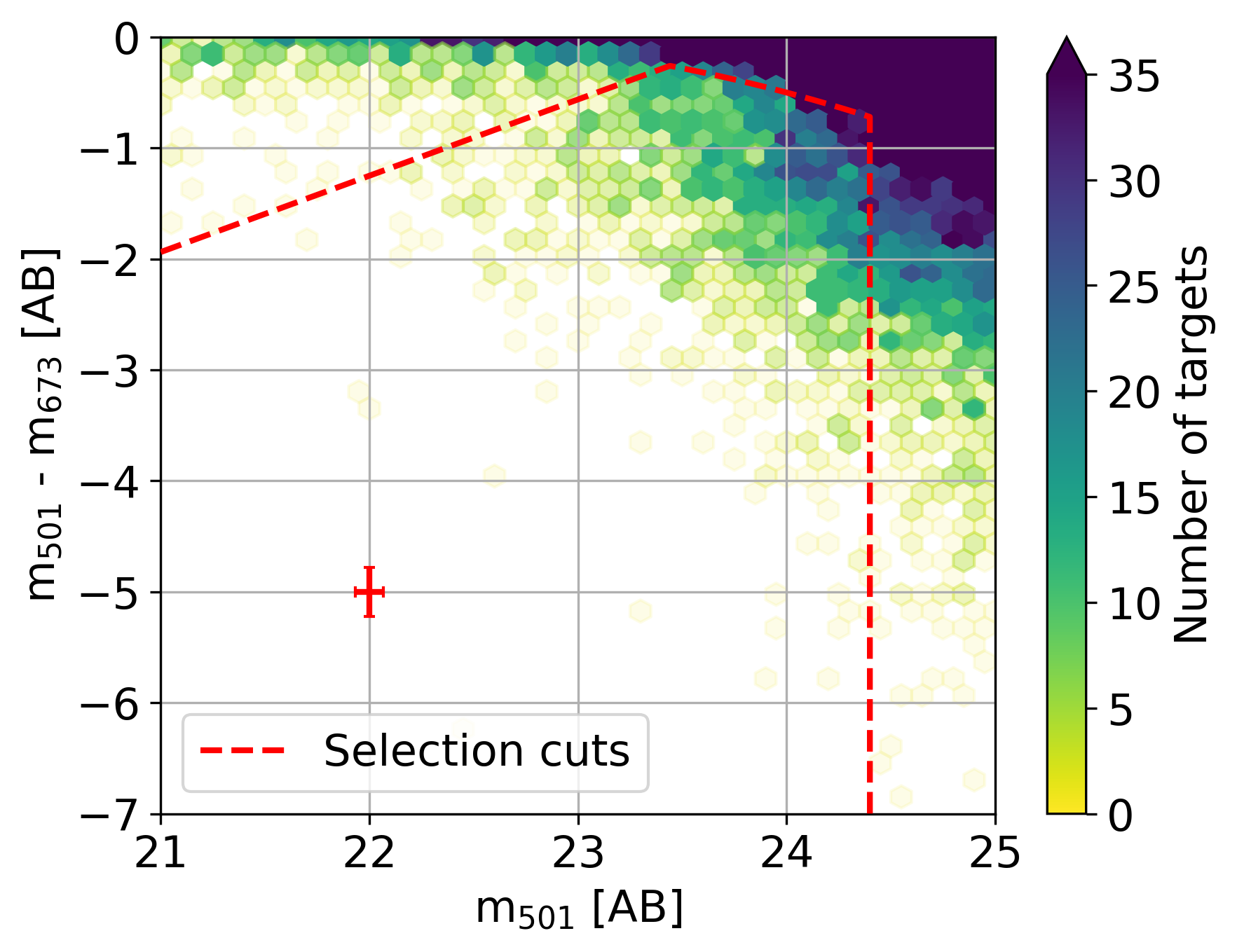}\\
    \includegraphics[width=0.45\columnwidth]{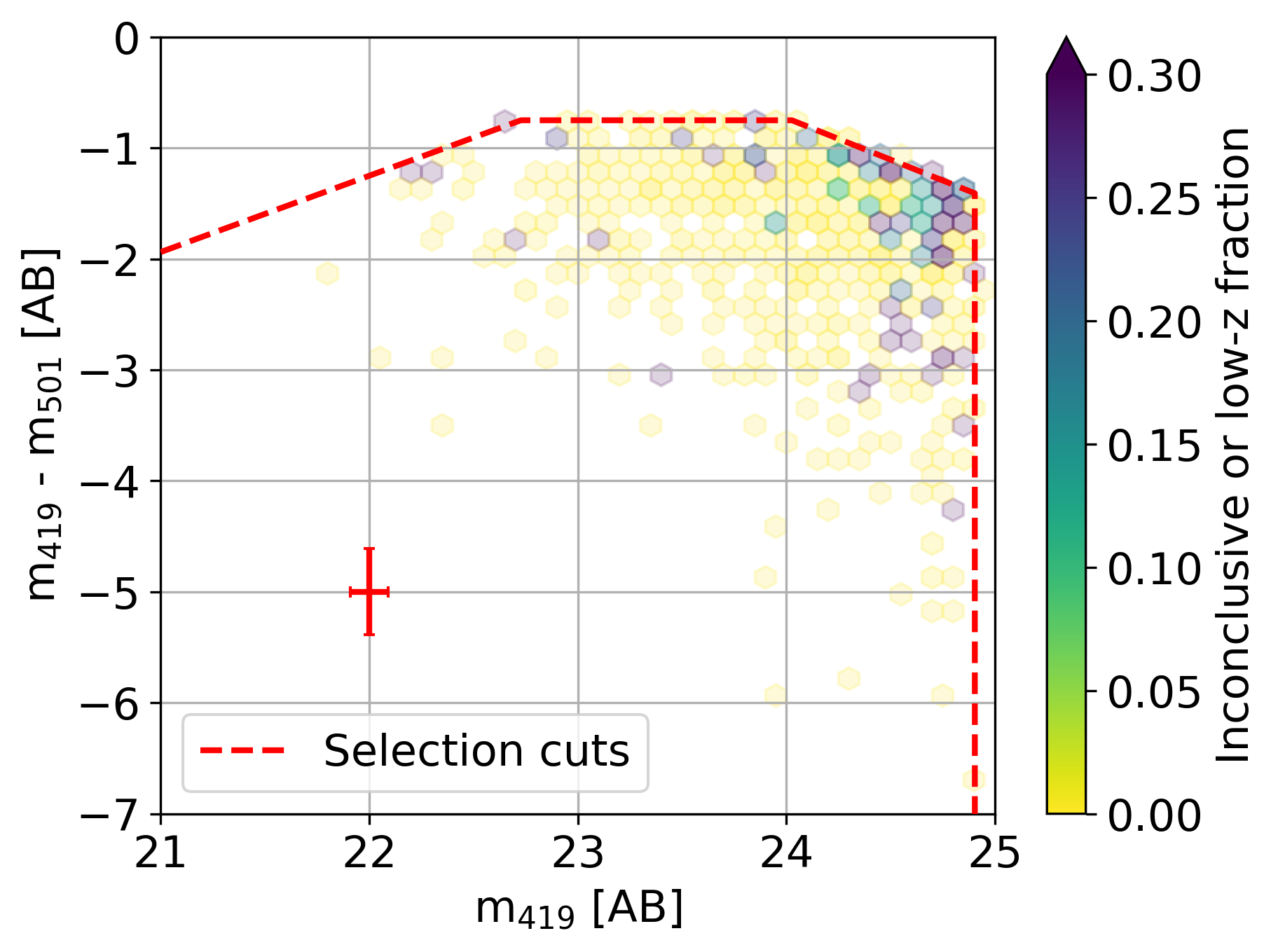} &
    \includegraphics[width=0.45\columnwidth]{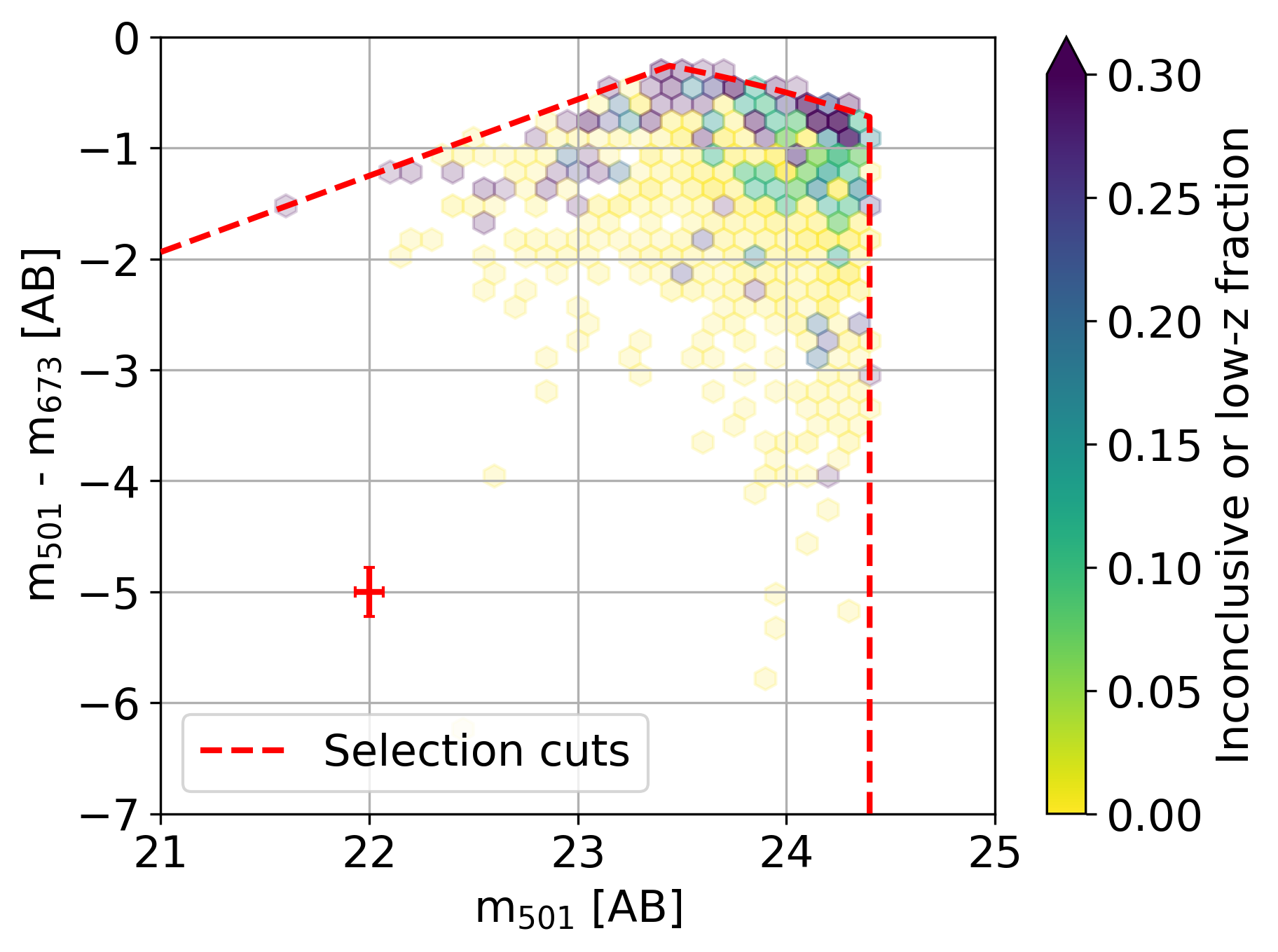}\\
\end{tabular}
\caption{Color-magnitude diagrams for the refined N419 ($z\simeq 2.45$; left) and N501 ($z\simeq 3.1$; right) samples showing the number of targets (upper panels; on a linear color scale) or the `contamination fraction' (lower panels).  The red error bars show the typical observational uncertainty, computed for the median values of the selections: (24.3, -1.6) for N419 and (24.00, -1.3) for N501.  See text for further details and the exact cuts adopted on the narrow bands.}
\label{fig:cuts}
\end{figure}

Using the final lists of high-quality redshifts, we optimized the selection criteria to result in high purity samples with a well-measured interloper fraction.  We shall refer to this as the ``refined'' sample below, and caution that it has a complex relationship to a sample selected only on rest-frame equivalent width or one which has been selected primarily for galaxy-evolution studies.  This reflects the optimization for a different purpose -- to enable cosmological studies we desire targets with a high probability of being at high redshift and that are likely to give good redshifts with modest cost.  To interpret the clustering measurements we desire a well-controlled interloper fraction.  The chosen selections are reported below, and the ODIN color-magnitude diagrams are displayed in Figure~\ref{fig:cuts}.  Further details are provided in \cite{odin_sample_selection}.

The adopted selection criteria for the $z\simeq2.45$ sample are: 
\begin{subequations}
    \begin{align}
		19 < \mfon & < 24.906 \label{eq:cut419_a}\\
		\mfon - \mfoo & < -0.75 \label{eq:cut419_b}\\
            \mfon - \mfoo & < -16.375 + 0.6875 \, \mfon \label{eq:cut419_d}\\
            \mfon - \mfoo & < 17.27 - 0.75 \, \mfon \label{eq:cut419_e}
	\end{align}
\end{subequations}
%
The adopted selection criteria for the $z\simeq3.1$ sample are:
\begin{subequations}
    \begin{align}
		18 < \mfoo < 24.40 \label{eq:cut501_a}\\
            \mfoo - \msst & < -60.5 + 5.5 \, \mfoo -0.125 \, \mfoo^2 \label{eq:cut501_c}\\
            \mfoo - \msst & < -16.375 + 0.6875 \, \mfoo \label{eq:cut501_d}
	\end{align}
\end{subequations}

\begin{figure}
\centering
\begin{tabular}{cc}
    \includegraphics[width=0.45\columnwidth]{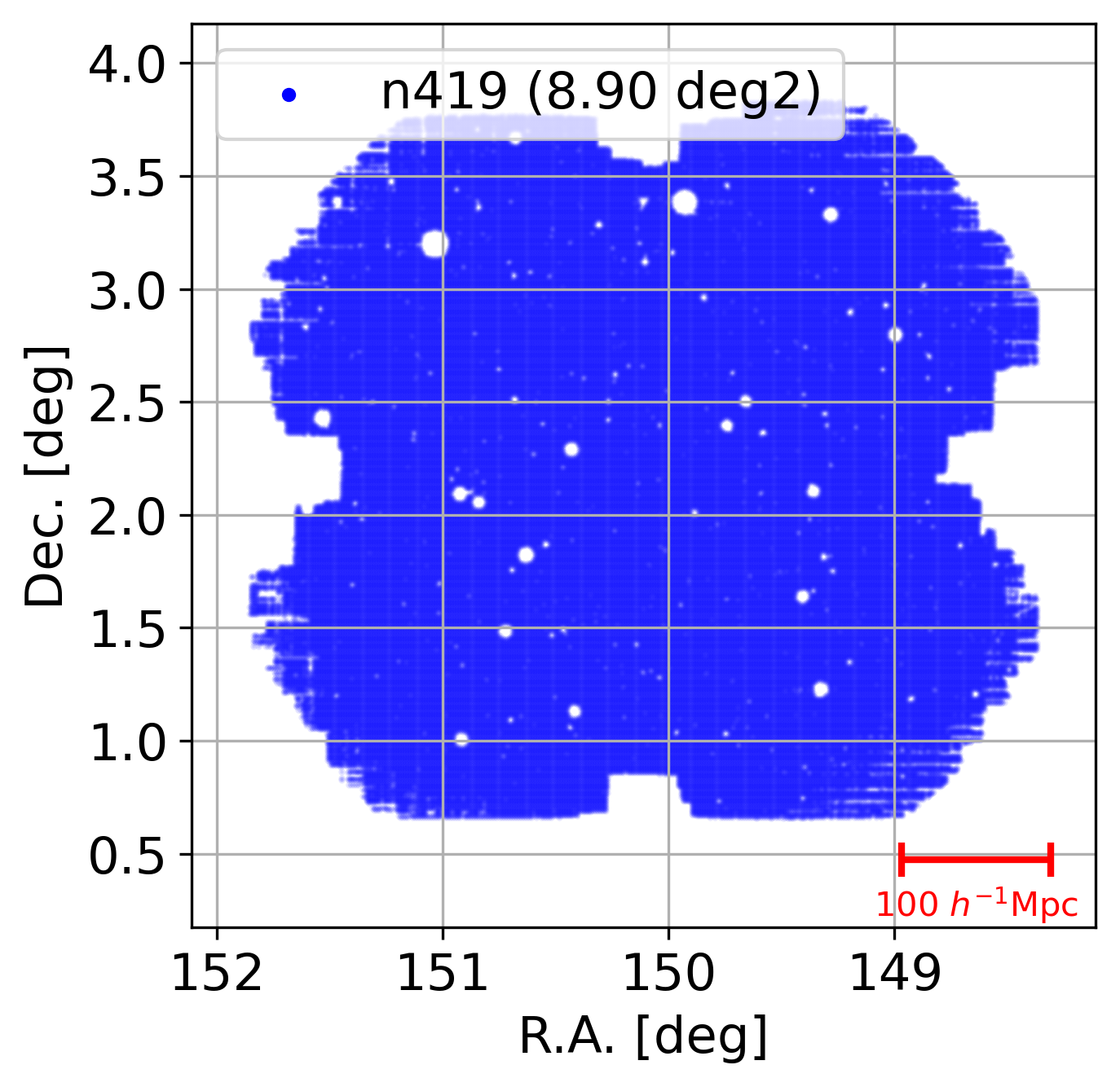} &
    \includegraphics[width=0.45\columnwidth]{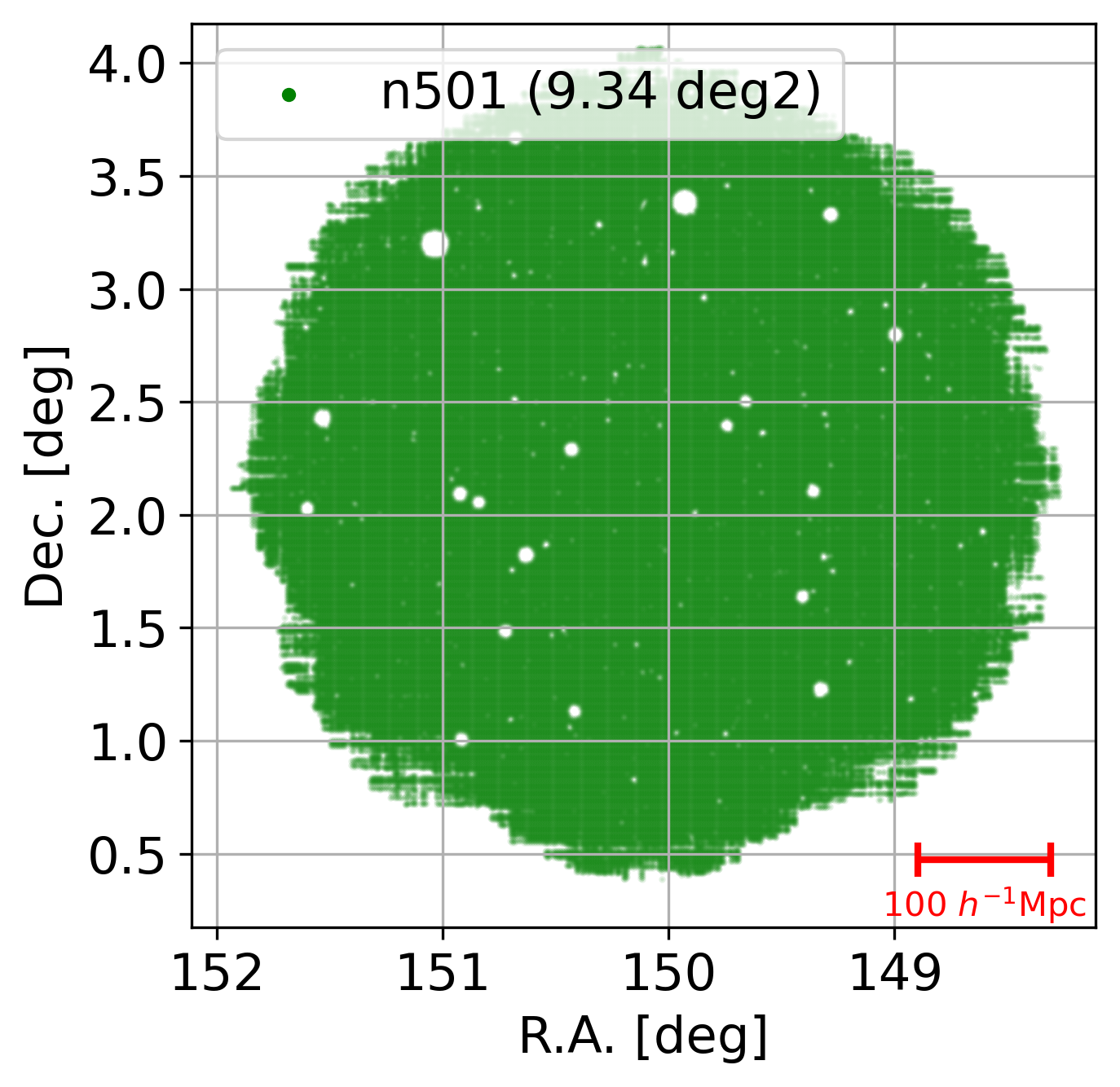}\\
\end{tabular}
\caption{Survey geometry for our analysis for each ODIN band.  The blue and green regions indicate the footprint, as determined from our random catalog, with the small (white) holes due to bright stars or other masksed regions.  Note the footprint for N419 differs from that for N501 due to a further restriction from the HSC imaging coverage (see text).  The inset bar shows $100\,h^{-1}$Mpc, comoving at the mean redshifts of the sample ($z\simeq 2.45$ and $3.1$ for the left and right panels, respecitvely).}
\label{fig:mask}
\end{figure}

We present in Table~\ref{tab:data} the number of candidates ($N_{\rm targ}$) and the number of observed candidates with more than 1.5 hrs of effective time ($N_{\rm obs}$). Among those, $N_{\rm VI,ok}$ is the number having a VI-validated redshift and $N_{\rm VI,zok}$ is the number of those with $2.35 < z_{\rm VI,ok} < 2.50$ for N419 and $3.075 < z_{\rm VI,ok} < 3.175$ for N501.

By definition, we do not know the redshift of the spectra with a non-conclusive VI. We thus provide two values of the interloper fraction, which should bracket the truth.  An optimistic estimation is to assume that the spectra with a non-conclusive VI have a similar redshift distribution to the ones with a conclusive VI: the interloper fraction would then be $f_{\rm int}=1 - N_{\rm VI,zok}/N_{\rm VI,ok}$.  A pessimistic estimate assumes that all the spectra with a non-conclusive VI are interlopers: the interloper fraction would then be $1-N_{\rm VI,zok}/N_{\rm VI}$.  From the VI and the pattern of failures, with the interloper and high-$z$ fractions being steady with exposure time but the redshift failures depending strongly on exposure time \cite{odin_sample_selection}, it appears the optimistic scenario is more likely for the N419 sample for which we thus take a fiducial $f_{\rm int}=0.02$.  The N501 sample is more ambiguous and we take $f_{\rm int}=0.08$ as our fiducial value.  Variations over the full range $f_{\rm int}=0.04-0.11$ lead to changes in the amplitude of the inferred clustering of $-8\%$ or $+7\%$ of the fiducial.  This translates into a $3-4\%$ systematic uncertainty in the clustering length ($r_0$), which is smaller than our statistical errors. We shall neglect this henceforth, pending further spectroscopic follow-up (though we list this as a further systematic uncertainty in the measured clustering amplitude in Table \ref{tab:samples}).

\begin{figure}
\centering
\resizebox{\columnwidth}{!}{\includegraphics{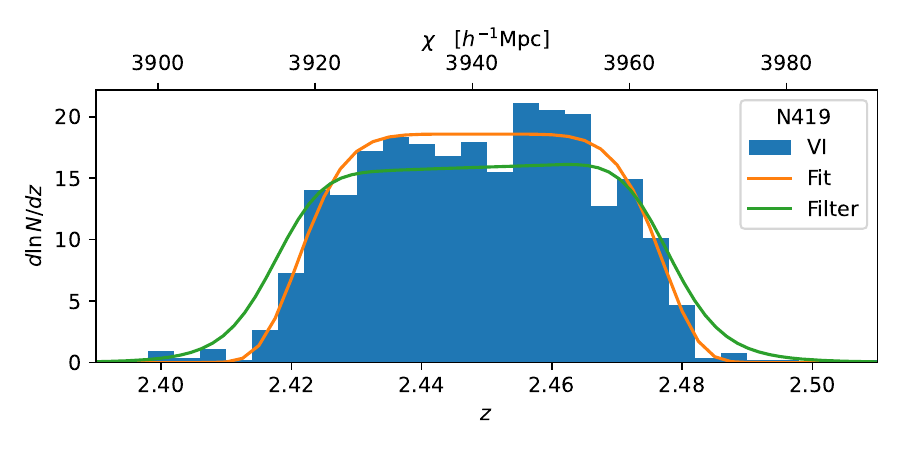}}
\resizebox{\columnwidth}{!}{\includegraphics{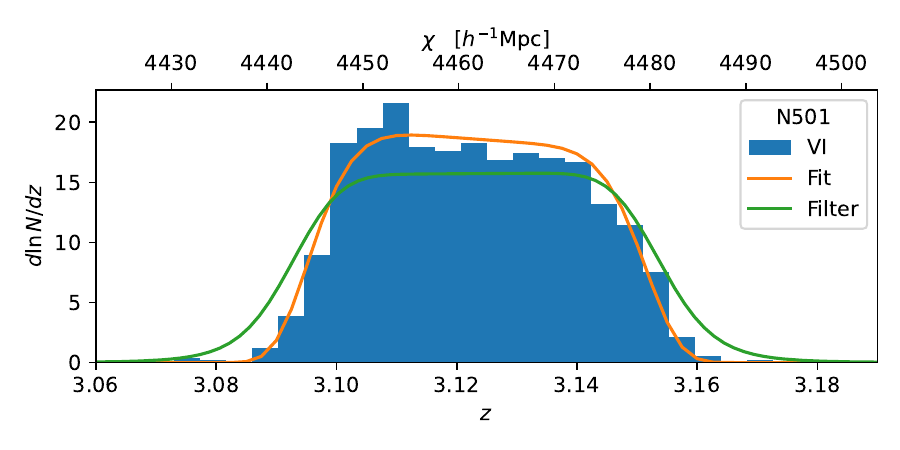}}
\caption{The (normalized) redshift distributions for the N419 ($z\simeq 2.45$) and N501 ($z\simeq 3.1$) samples described in the text.  The blue histograms show the `secure' redshifts of the visually inspected sources (VI), while the orange line is a simple analytic fit used in the theoretical model and the green line shows the filter transmission curve converted to redshift at the Ly$\alpha$ wavelength.  On each panel the lower $x$-axis shows the redshift while the upper $x$-axis shows the corresponding comoving, radial distance (in $h^{-1}$Mpc) assuming our fiducial cosmology \cite{Planck18-I}.}
\label{fig:dndz}
\end{figure}

\subsection{Survey geometry}

The survey geometry is defined by the coverage of the narrow- and broad-band imaging, along with the angular masking done in the target selection. The considered geometry for each band is displayed in Figure~\ref{fig:mask}; we describe below how we construct it.

First, we generate $10^6$ randoms per deg$^{2}$ uniformly within a 1.9 deg radius from our COSMOS field center (R.A., Dec.) = ($150.11^\circ$, $2.173^\circ$).  We use the approach described in ref.~\cite{Myers23}, adapted to our custom Tractor products.  In particular, in the {\tt MASKBITS} content, the three ODIN bands (N419, N501, N673) replace the $g$, $r$, and $z$ broad bands of the Legacy Surveys.
This allows us to propagate the angular quantities used in the target selection, and apply those cuts on the randoms:
(1) ${\tt NOBS \ge 10}$ requiring at least ten images at the central pixel location for each of the two ODIN bands used to build the ODIN color in the target selection;
(2) ${\tt MASKBITS} = 0$, which discards unreliable (e.g.\ saturated) pixels of the imaging, along with regions masked by the {\tt BRIGHT} and {\tt MEDIUM} Legacy Surveys stellar masks;
(3) an additional stellar mask rejecting objects within a radius
\begin{equation}
    R = 0.07^\circ \left(\frac{6.3}{\mathtt{GAIA\_PHOT\_G\_MEAN\_MAG}}\right)^2 
\end{equation}
of a star brighter than 9 mag.
For the N419 selection, we add a mask mapping the HSC imaging coverage.  Since the N501 selection is based on a ``LS or HSC'' set of cuts and the union of the LS and HSC images fully covers our $1.9^\circ$-radius region, there is no need for an extra mask.  Finally, based on those cut randoms, we build a high-resolution {\sc HealPix} pixel map \cite{Gorski05} at $\mathrm{Nside}=8192$ which can apply to mocks in the clustering analysis.

\section{Clustering analysis}
\label{sec:clustering}

\begin{table}[t]
\centering
\begin{tabular}{lccccccccc}
\hline
 Filter & Redshift      & $\mathrm{lg} F_{\rm Ly\alpha}^{\rm med}$ & $f_{\rm int}$ & $\chi_0$ & $\Delta$
        & $d\chi/dz$    & $r_0$   & $10^3\,\bar{n}$ & $b$    \\
\hline
{\bf N419} & $2.4\pm 0.03$ & -16.0 & 0.02 & 3941 & $50$ & 829 & $3.0\pm 0.2$ & 1.2 & $1.7\pm 0.2$ \\
{\bf N501} & $3.1\pm 0.03$ & -16.1 & 0.08 & 4448 & $38$ & 633 & $3.0\pm 0.2$ & 1.0 & $2.0\pm 0.2$ \\
\hline
\end{tabular}
\caption{Inferred properties of the refined LAE samples in each redshift range.  The median Ly$\alpha$ flux, $F_{\rm Ly\alpha}^{\rm med}$, is in erg~s$^{-1}$~cm$^{-2}$ and $\mathrm{lg}$ indicates $\log_{10}$. The interloper fraction, $f_{\rm int}$, is inferred from DESI spectroscopy.  The comoving distance to the center of the shell ($\chi_0$), the FWHM of the shell ($\Delta$), the distance-redshift slope, the correlation length, $r_0$, and the 3D number density are in $h^{-1}$Mpc units to match the conventions typically employed in large-scale structure.  The large-scale bias, $b$, is inferred from the mock catalogs (see text).  For the N501 sample, the full range of uncertainty in $f_{\rm int}$ would lead to an additional $\approx 3-4\%$ systematic error on $r_0$ or $b$.}
\label{tab:samples}
\end{table}

The DESI spectroscopy provides us with constraints on the redshift distribution and interloper fraction for the ODIN galaxies.  We see from Fig.~\ref{fig:dndz} that the ODIN galaxies lie in thin shells of widths of $\Delta z=0.06$ at $z=2.45$ and 3.1.  These translate into comoving distance `depths' of $\Delta\simeq 50$ and $38\,h^{-1}$Mpc (Table \ref{tab:samples}).  For such narrow slices in redshift the majority of the large-scale clustering information lies in the angular clustering.  While for such small sky areas the corrections for beyond-plane-parallel are small, the narrow depth of the ODIN survey means that we cannot neglect redshift-space distortions\footnote{For a top-hat selection of width $\Delta\chi$ the average line-of-sight separation is $\Delta\chi/3\sim 10\,h^{-1}$Mpc.  Thus on scales $\mathcal{O}(10\mathrm{Mpc})$ redshift-space distortions give a correction to the clustering of tens of percent.} (RSD).  Further, the Limber approximation \cite{Limber53}, which assumes a depth much larger than the transverse scale of interest, does not hold.  However because the clustering is not diluted by projection over a large depth, the magnification bias is very small.

Assuming $\phi(\chi)\propto H(z)\,dN/dz$ is the radial distribution as a function of comoving distance, normalized to $\int \phi\, d\chi=1$, the (angular) power spectrum of LAEs observed within the shell in the flat-sky approximation is
\begin{equation}
  C_\ell = \int_0^\infty\frac{dk_\parallel}{\pi \chi^2}
  \ P\left(k_\perp=\ell\chi^{-1},k_\parallel\right)
  \left| \widetilde{\phi}(k_\parallel)\right|^2
\end{equation}
with $\widetilde{\phi}$ the Fourier transform of $\phi$, $\chi$ the comoving distance to the shell center and $P(k_\perp,k_\parallel)$ is the 3D power spectrum.  Variation of $dN/dz$ over the field is negligible for our study \cite{Lizancos23}.  Note for shells of full-width $\Delta$ the line-of-sight window, $\widetilde{\phi}^2(k_\parallel)$, falls off for $k_\parallel\,\Delta\gg 1$ and so we expect the integral to depend on $k_\parallel < \mathcal{O}(\Delta^{-1})\sim 0.1\,h\,\mathrm{Mpc}^{-1}$.  If the shell is further divided into redshift slices with distributions $\phi_a$ then $|\widetilde{\phi}|^2$ is replaced by a product.  In the specific case of disjoint, top-hat distributions of width $\Delta$ whose centers are separated by $\chi_{ab}$ the line-of-sight window function becomes $\cos[k_\parallel\chi_{ab}]\, j_0^2(k_\parallel\Delta/2)$ with $j_0=\sin x/x$ the spherical Bessel function of order 0.

The angular correlation function can be written as
\begin{eqnarray}
  w(\theta) &=& \sum_\ell\frac{2\ell+1}{4\pi} C_\ell P_\ell(\cos\theta) 
  \simeq \int\frac{\ell\ d\ell}{2\pi} C_\ell J_0(\ell\tilde{\omega}) \nonumber \\
  &=& \int \frac{d^3k}{(2\pi)^3}
  \ P\left(k_\perp,k_\parallel\right)
  \ \widetilde{\phi}^2(k_\parallel) J_0(k_\perp\chi\ \tilde{\omega})
\end{eqnarray}
with $J_0$ the cylindrical Bessel function of order $0$ and $\tilde{\omega}=2\sin(\theta/2)$. Note this has the expected form of an integral of the power spectrum multiplied by a line-of-sight window function ($\widetilde{\phi}^2$) and a transverse window function ($J_0$).  This can also be rewritten directly in terms of the 3D correlation function as
\begin{equation}
  w_\theta(R) = 2\int_{0}^{\infty} dy\ W(y)\,\xi(\sqrt{R^2+y^2},\mu_y)
  \quad , \quad
  W(y) =  \int_0^\infty d\bar{\chi}\ \phi(\bar{\chi}-y/2)\phi(\bar{\chi}+y/2)
\label{eqn:wR_xi}
\end{equation}
where $\mu_y=y/\sqrt{R^2+y^2}$ and we have converted angles to (transverse) distances with $R=\chi\theta$, denoting this statistic $w_\theta(R)$.  In the limit of a top-hat redshift distribution of full-width $\Delta$ this further simplifies to
\begin{equation}
  w_\theta(R) = 2\int_0^1 dY\ (1-Y)\,\xi(\sqrt{R^2+Y^2\Delta^2},\mu)
  \qquad (y=Y\Delta)
\end{equation}
The integral can be done with simple quadrature given a model for $\xi(s,\mu)$, and the case of power-law power spectra or correlation functions is treated in Appendix \ref{app:power-law}.  The generalization for the cross-spectrum of two shells is straightforward.

{}From the data and random catalogs we compute the angular power spectrum (using the method of ref.~\cite{Lizancos24}) and correlation functions, $w_\theta(R)$ (using the Landy-Szalay estimator \cite{Landy93}).  The field is large enough that the integral correction in $w_\theta(R)$ can be neglected on the scales of interest, which we have verified with our N-body-based mock catalogs.  Angles are converted to distances assuming the median redshift ($\chi_0$; listed in Table \ref{tab:samples}).  Since the non-LAE targets are physically well separated (in redshift) from the LAEs the measured clustering can be well approximated as $w_\theta\simeq (1-f_{\rm int})^2 w_{\rm LAE} + f_{\rm int}^2 w_{\rm int}$, where $f_{\rm int}$ is the interloper fraction.  Neglecting the clustering of the interlopers, since $f_{\rm int}$ is small and the interlopers appear to be spread over redshift, we can compute the LAE clustering by dividing our measured $w_\theta(R)$ by $(1-f_{\rm int})^2$.  The plotted correlation functions have had this correction applied, assuming the fiducial value of $f_{\rm int}$ quoted in Table \ref{tab:samples}.

\section{Mock catalogs}
\label{sec:mocks}

During the planning of the observations and for the analysis we have used a set of mock catalogs.  These mock catalogs are built upon halo catalogs from the {\sc AbacusSummit} cosmological N-body simulation suite \cite{Maksimova21}, produced with the {\sc Abacus} N-body code \cite{Garrison21}.  Our fiducial simulation was of the $\Lambda$CDM family employing $6300^3$ dark matter particles in a box of side length $1\,h^{-1}$Gpc ($m_{\rm part}=3.5\times 10^8\,h^{-1}M_\odot$; see ref.~\cite{Maksimova21} for more details).  As well as the simulations released in ref.~\cite{Maksimova21}, we use two additional simulations that were run on the Perlmutter supercomputer at the National Energy Research Scientific Computing center.  The first simulation utilized $6000^3$ particles in a $1.25\,h^{-1}$Gpc box ($m_{\rm part}=7.9\times 10^8\,h^{-1}M_\odot$) with outputs from $z=5$ to $z=2$.  The force softening was held fixed in proper coordinates, reaching $0.025$ of the mean interparticle spacing at $z=2$.  The second utilized $6912^3$ particles in a $750\,h^{-1}$Mpc box ($m_{\rm part}=1.1\times 10^8\,h^{-1}M_\odot$), also with outputs down to $z=2$.  We used these additional simulations to confirm that our results are converged with respect to halo mass resolution and finite volume effects.
In all cases, the halos are populated with mock galaxies using the AbacusUtils\footnote{https://abacusutils.readthedocs.io/en/latest/} software.

The halo occupation distribution of LAEs is quite uncertain, so our modeling is at best approximate.  Recent simulations \cite{McQuinn07,Nagamine10,Yajima12,Lake15,Garel15,Weinberger18,Weinberger19,Ravi24} and observations \cite{Bielby16,Kusakabe18,Khostovan19,Ouchi20} suggest that LAEs occupy a fraction of the low-mass halo population with the number of galaxies growing more slowly than halo mass towards large masses.  We build such catalogs from the AbacusUtils software, which implements \cite{Yuan22} a standard halo occupation distribution modeling \cite{Asgari23} with occupancy following the form introduced by ref.~\cite{Zheng07}.  Each halo, of mass $M_h$, has a central galaxy selected from a binomial distribution and a Poisson-distributed number of satellites with respective means
\begin{equation}
  \bar{n}_{\rm cen}(M_h) = \frac{1}{2}\mathrm{erfc}\left[ \frac{\ln M_{\rm cut}/M_h}{\sqrt{2}\,\sigma} \right]
  \quad , \quad
  \bar{n}_{\rm sat}(M_h) = \left[ \frac{M_h-\kappa M_{\rm cut}}{M_1} \right]^\alpha
  \bar{n}_{\rm cen}(M_h)
\end{equation}
and we take $\alpha<1$ and apply a random downsampling of the galaxies to match the observed number density \cite{Gawiser07,Kovac07}.  This allows fractional occupancy even at low halo mass and has LAEs preferentially avoiding high mass halos\footnote{The exact behavior of the HOD for high masses is not too important for our purposes.  Our LAEs have $b>1$ and so populate halos on the exponentially declining part of the halo mass function.  Halos significantly more massive than $M_{\rm cut}$ are very rare, and thus our results are quite insensitive to the behavior of $\bar{n}_{\rm cen}$ or $\bar{n}_{\rm sat}$ for these masses.}.  In the above, $M_{\rm cut}$ parameterizes the mass threshold below which halos do not host LAEs, with a threshold sharpness set by $\sigma$.  The value of $M_1$ indicates where halos host $\approx 1$ satellite in addition to a central, with $\kappa$ setting the low-mass suppression of the satellite occupancy. Ref.~\cite{Ravi24} show that such a form matches star-forming galaxies selected in cosmological hydrodynamic simulations.  Physically such occupancy arises due to selections on narrow and broad band color for which only a fraction of the parent population passes (due to a combination of the fraction of time spent in an appropriately low-dust burst of star formation and any orientation effects that influence the ``opening angle'' over which Lya photons escape after radiative transfer through the ISM) and on Ly$\alpha$ flux and on equivalent width \cite{Dijkstra10}, for which our procedure amounts to a random selection near threshold.  The central galaxy is then assigned to the center of mass of the halo, with the velocity vector also set to that of the center of mass of the halo. Satellite galaxies are assigned to particles of the halo with equal weights.  Following the preliminary investigations reported in \cite{Ravi24}, and in the absence of observational evidence to the contrary, we do not include secondary/assembly bias or velocity bias in the model \cite{Yuan22}.

Since the number density provides only an upper limit on $M_{\rm cut}$, the information on the characteristic mass of LAE-hosting halos is mostly provided by the large-scale bias.  The scale-dependence of the clustering could in principle provide constraints on the detailed halo occupation, but with current error bars the problem is underconstrained.  Rather than attempt a detailed exploration of the parameter space, we have selected from a grid of HOD models a subset that provide good fits to our data ($\chi^2<18$ for 10 d.o.f.).  These provide us nearly noise-free `observations' and a qualitative sense of the model uncertainty and have been made publicly available (see \S\ref{sec:data_availability}).

The mock catalogs are generated from single snapshots of the $N$-body simulation at fixed redshift, using the closest redshift to the observed galaxies (i.e.,~$2.5$ and $3.0$ respectively).
Real- and redshift-space correlation functions and power spectra are computed directly from the (periodic box) outputs for each model, using the technique of Zeldovich control variates \cite{Kokron22,DeRose23,Hadzhiyska23} to reduce the sampling fluctuations at large scales.  In addition, mock observations are performed by projecting the galaxies onto the sky.  By applying random offsets parallel to each of the box axes we generate multiple independent volumes.  For each volume we apply the same masks as for the real data and randomly downsample the objects to match the observed number density and redshift distribution.  This ensures the clustering amplitude, shotnoise and projection of distances to angles matches that in the data.  We compute the covariance of the clustering, with the same binning and statistics as used in the real data, for 1024 mocks in order to generate a Monte-Carlo covariance matrix.  Given the small number of bins that we measure this number is more than sufficient to give a converged covariance matrix.

\begin{figure}
    \centering
    \resizebox{\columnwidth}{!}{\includegraphics{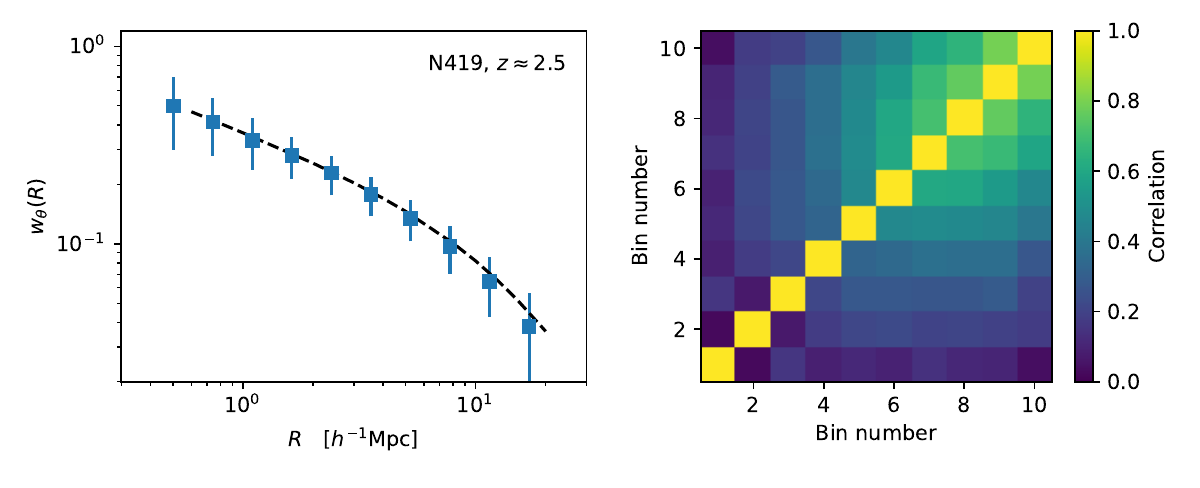}}
    \resizebox{\columnwidth}{!}{\includegraphics{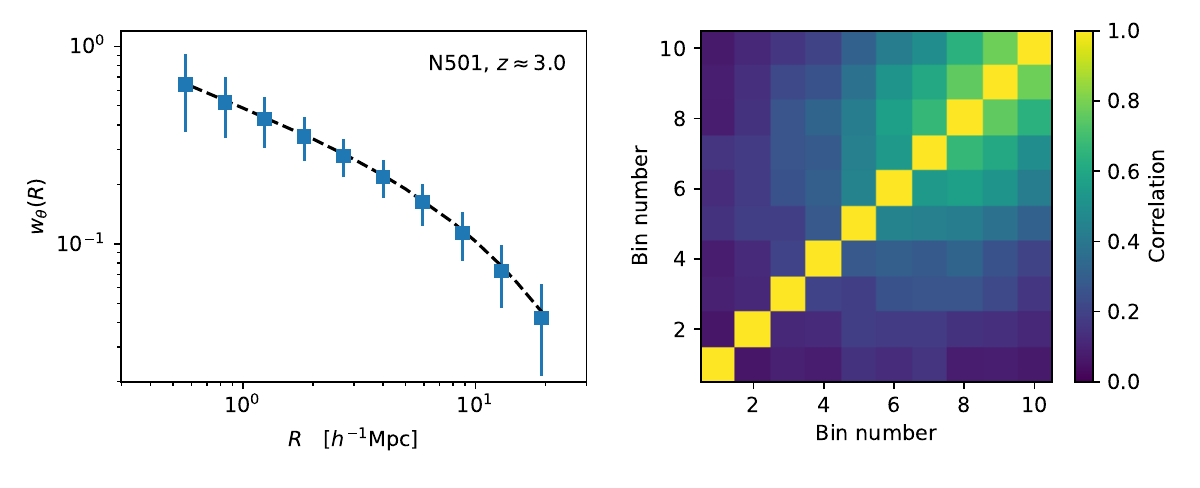}}
    \caption{(Left) The average angular clustering measured from the 1024 mock catalogs for the N419 and N501 samples in the COSMOS field using the same pipeline as the observations, compared to the prediction from the correlation function multipoles measured directly in the N-body simulation using Eq.~(\ref{eqn:wR_xi}).  The agreement indicates the pipeline returns unbiased answers.  The errors on the points indicate the standard deviation measured from the mocks, as appropriate for a single field, i.e.\ not the error on the mean.  (Right) The correlation matrix.  Note that at small scales the bins are close to independent, as expected for shot-noise, while at larger scales neighboring bins become more correlated as expected in the sample variance dominated regime. }
    \label{fig:mock_test}
\end{figure}

While we have set the mean number of objects per field to match the observed angular number density, the actual number of mock LAEs per field fluctuates by $\approx 10\%$ from realization to realization.  This is due to sample variance from the small observational volume of a single field.  We make no attempt to correct for this effect.  The left panel of Fig.~\ref{fig:mock_test} shows the errors on the clustering inferred for a single field (i.e., not the error on the mean).  We see that our pipeline returns an unbiased measure of the clustering from the mocks and a high-precision measurement of the clustering is anticipated from the data.  At small scales the bins are largely uncorrelated (right panel of Fig.~\ref{fig:mock_test}) as expected for shot-noise.  As we move to larger scales the points become increasingly correlated as we enter the sample-variance dominated regime and the finite survey volume becomes important.  We use the full covariance matrix for all of our inferences, thus including these correlations.

Since the main purpose of our simulations is to provide mock catalogs that match the clustering of the samples under consideration, we regard them as adequate.  However future work in this direction is clearly desirable.  A preliminary investigation of the occupancy of LAEs within the MTNG simulation \cite{Ravi24} suggests that our modeling is reasonable under the approximation that Ly$\alpha$ line strength closely follows the star-formation rate (which is what the simulation determines best), but additional investigation of the LAE population in hydrodynamic simulations including modeling of the observed Ly$\alpha$ line flux would obviously be valuable \cite{McQuinn07,Nagamine10,Yajima12,Lake15,Garel15,Weinberger18,Weinberger19}.

\section{Results}
\label{sec:results}

\begin{figure}
    \centering
    \resizebox{0.475\columnwidth}{!}{\includegraphics{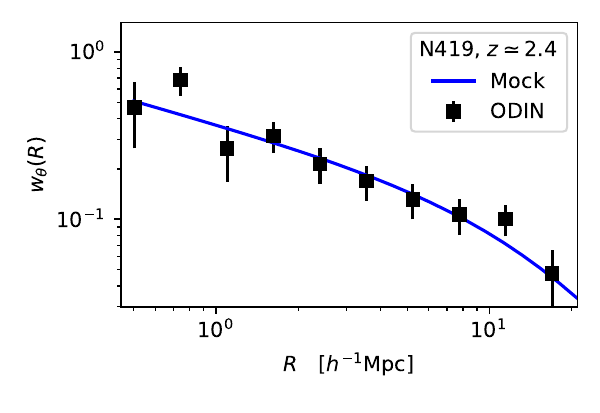}}
    \resizebox{0.475\columnwidth}{!}{\includegraphics{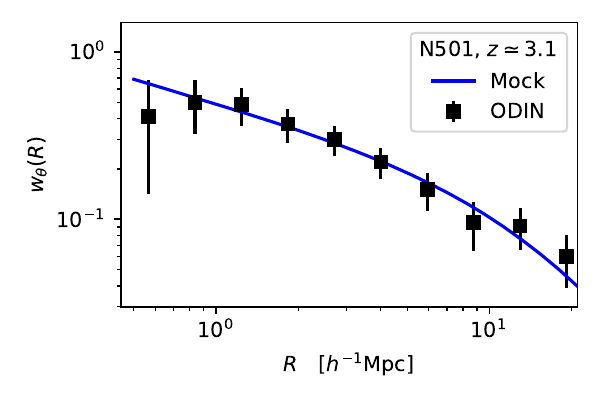}}
    \caption{Clustering predicted in the fiducial mock catalogs (lines) compared to the ODIN data (black points with error bars) for the N419 and N501 samples.  The error bars indicate the diagonal entries of the covariance matrix derived from the same mock catalogs and described further in the text. }
    \label{fig:wR_HOD}
\end{figure}

\begin{figure}
    \centering
    \resizebox{0.475\columnwidth}{!}{\includegraphics{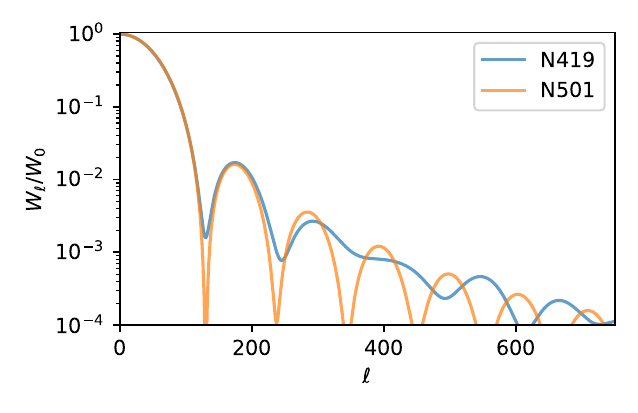}}
    \resizebox{0.475\columnwidth}{!}{\includegraphics{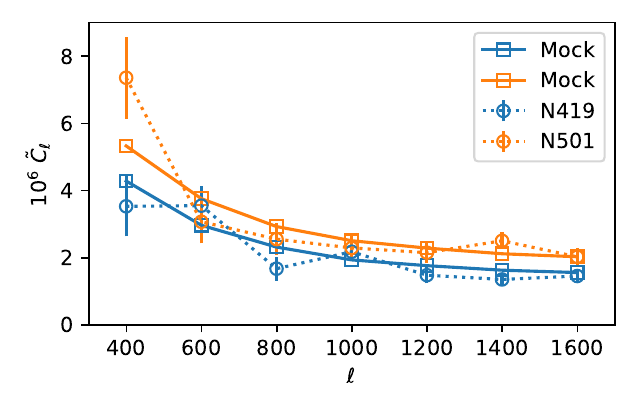}}
    \caption{Window function (left) and angular power spectra (right) for the N419 and N501 samples in the COSMOS field.  The small angular extent of the survey means that multipoles are coupled, with an extent determined by $W_\ell$.  For this reason we bin the power spectrum in bins of $\Delta\ell=200$.  The circles connected by dashed lines indicate the pseudo-$C_\ell$ measured on the data while the squares with error bars indicate the mean and standard deviation from the mocks described in the text (which were fit to the $w_\theta(R)$ measurements of Fig.~\ref{fig:wR_HOD}).  The errors are correlated at the $\approx 20\%$ level, and the spectra are shot-noise dominated beyond $\ell=10^3$.}
    \label{fig:power_spectra}
\end{figure}

\begin{figure}
    \centering
    \resizebox{0.475\columnwidth}{!}{\includegraphics{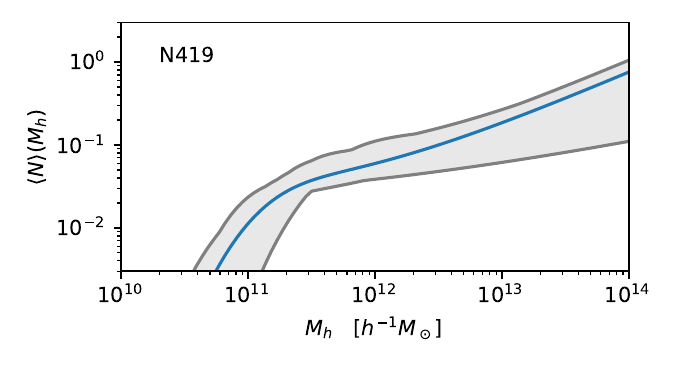}}
    \resizebox{0.475\columnwidth}{!}{\includegraphics{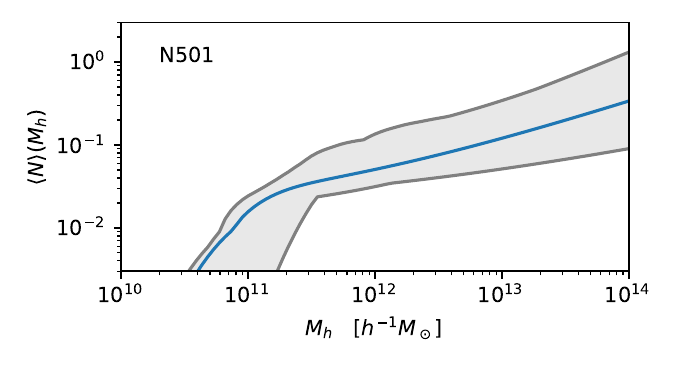}}
    \caption{Mean LAE halo occupation in the mock catalogs.  The blue lines show the best-fitting model in our grid, the grey bands show the range of $\langle N_{\rm gal}\rangle$ in the HOD models that provide acceptable fits to the ODIN data ($\chi^2<18$ for 10 d.o.f.) for the N419 (left) and N501 (right) samples.  As discussed in the text, the high-$M_h$ behavior is not well constrained by the data but this presents little difficultly for cosmological interpretation. }
    \label{fig:HOD}
\end{figure}

Fig.~\ref{fig:wR_HOD} shows the measured configuration-space clustering, $w_\theta(R)$, with angles converted to tranverse distances using the $\chi_0$ listed in Table \ref{tab:samples}, compared to the predictions of our mock catalogs.  The error bars are determined from the mock catalog Monte-Carlo observations and represent the square root of the diagonal of the covariance matrix (see Fig.~\ref{fig:mock_test}).  Fig.~\ref{fig:power_spectra} shows the angular power spectra, and the window function, for the samples in the COSMOS field.  Again the error bars represent the square root of the diagonal of the covariance matrix.  The lines indicating the mock catalog angular power spectra are for the model fit to $w_\theta(R)$, i.e.,~the same model shown in Fig.~\ref{fig:wR_HOD}, and not a separate fit to the angular power spectrum data itself.
For completeness, Fig.~\ref{fig:HOD} shows the mean halo occupation, $\langle N_{\rm gal}(M_h)\rangle$, of the HOD models that provide acceptable fits to the $w_\theta(R)$ data -- the parameters and $w_\theta(R)$ predictions for the entire grid as well as the best-fit model are publicly available (see \S\ref{sec:data_availability}).

\begin{figure}
    \centering
    \resizebox{\columnwidth}{!}{\includegraphics{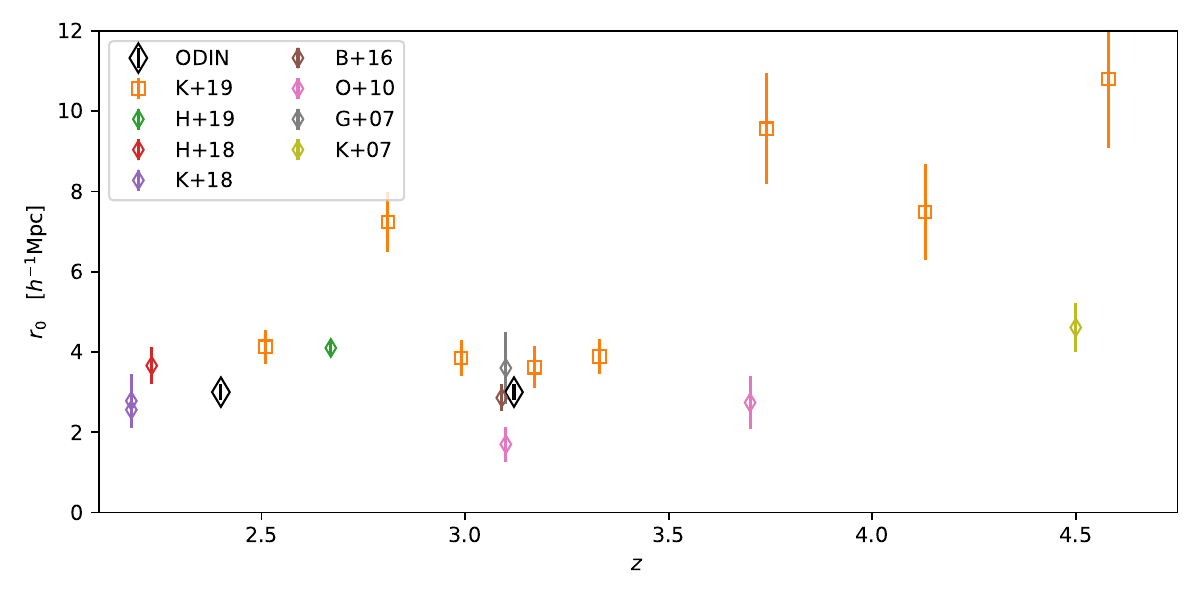}}
    \caption{Comparison of our correlation length estimates to a selection of earlier work \cite{Khostovan19,Hong19,Hao18,Kusakabe18,Bielby16,Ouchi10,Gawiser07,Kovac07}.  Our results are generally consistent with earlier estimates, despite differences in selection.  At 9 sq.deg.\ our results represent the largest contiguous area amongst these surveys.}
    \label{fig:lae_r0}
\end{figure}

The clustering is relatively well described by a power-law for $1<R<10\,h^{-1}$Mpc, and it has been common in the literature to fit a power-law to the data (see also Appendix \ref{app:power-law}).  Several further approximations need to be made in order to fit a power-law model, most notably how to handle redshift-space distortions.  The size of the distortion depends upon scale and bias, with the largest effect arising when $R\sim\Delta$ and the bias is small.  To handle this we infer $r_0$ from the projected clustering, $w_p(R)$, measured in our mock catalogs.  Specifically for each HOD model we derive the goodness-of-fit to the data ($\chi^2$) and also the projected clustering, $w_p(R)$.  From $w_p(R_{\rm fid}=5\,h^{-1}\mathrm{Mpc})$ we infer $r_0$ assuming $\xi_{\rm real}(r)=(r_0/r)^{\gamma}$ with $\gamma=1.8$, matching the assumptions most commonly employed in the literature:
\begin{equation}
    r_0 = \left[ \frac{\Gamma(\gamma/2)\ w_p(R_{\rm fid})}{R_{\rm fid}^{1-\gamma}\sqrt{\pi}\,\Gamma([\gamma-1]/2)} \right]^{1/\gamma}
    \quad .
\end{equation}
We find a good correlation between $r_0$ and $\chi^2$, suggesting that this clustering amplitude is a good summary statistic for the HOD models we investigate (see also ref.~\cite{Ebina24}).  We then compute a best fit and standard deviation from the $\chi^2$ minimum and $\Delta\chi^2$ assuming Gaussian statistics.  These values are reported in Table \ref{tab:samples} and compared to earlier estimates of the same quantity in Fig.~\ref{fig:lae_r0}.  We find that our measured clustering is lower than the predictions for similar samples in the simulation of ref.~\cite{Ravi24}, though within the envelope of previous observational determinations even though our selection is slightly different.

A key quantity of interest for cosmological applications of LAEs is the large-scale bias, which impacts the degree of anisotropy from redshift-space distortions and in combination with the 3D number density the `signal to noise' ratio for the power spectrum.  The uncertainties from the ODIN observations become large on quasi-linear scales where we expect the bias to approach its scale-independent, large-scale value.  However we can use the mock catalogs to measure the bias of the models that provide good fits to the ODIN data over a wider range of scales by directly comparing the mock galaxy clustering to that of the underlying matter field in the simulations -- to the extent that these models describe the halo occupancy of LAEs in the real Universe this can give a guide to the degree of scale-dependence we expect.  We discuss this further below.

\section{Forecasts}
\label{sec:forecasts}

Since they are numerous, have a bias similar to that of massive galaxies at $z<1$ and prominent emission lines for ease of redshifting, LAEs make appealing targets for future redshift surveys aimed at high-$z$ large-scale structure and primordial physics \cite{Wilson19,Sailer21,Ferraro22,Ebina24}.  In this section we look at the implications of our observations, interpreted within the context of the mock catalogs we have produced.  While they are only our first attempt at modeling the LAE population, since in those catalogs we have access to the underlying dark matter, have no observational artifacts and almost no sample variance (due to our use of control variates, discussed above) they allow us to make inferences about how well future surveys could perform.  The surveys, and analytic theory, also provide a plausible extension of the observed clustering to larger scales where the cosmological signals largely lie.  For brevity we focus on the $z\simeq 3.1$ population here, though the results for $z\simeq 2.45$ are qualitatively similar.

\subsection{Real-space clustering}

\begin{figure}
    \centering
    \resizebox{0.485\columnwidth}{!}{\includegraphics{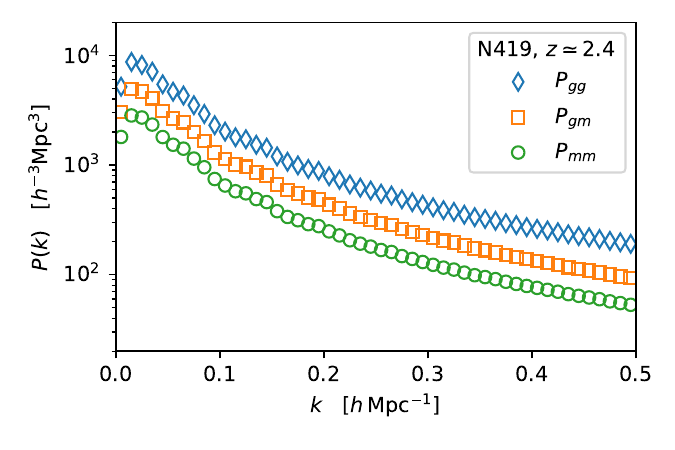}}
    \resizebox{0.485\columnwidth}{!}{\includegraphics{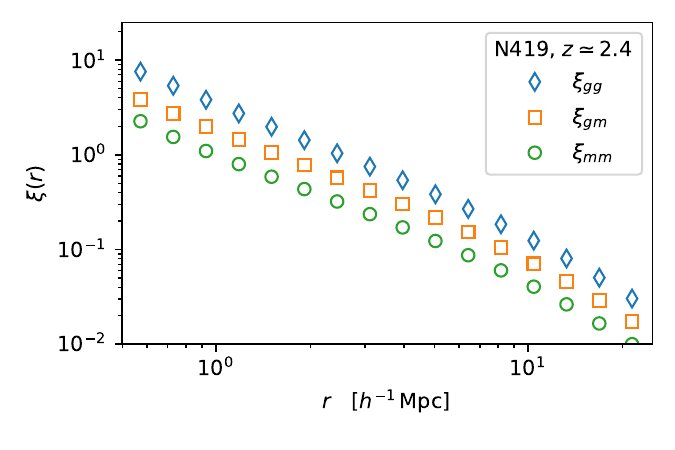}}
    \resizebox{0.485\columnwidth}{!}{\includegraphics{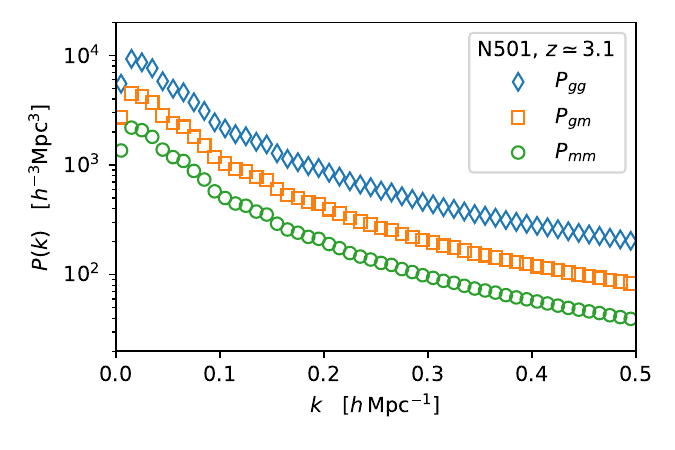}}
    \resizebox{0.485\columnwidth}{!}{\includegraphics{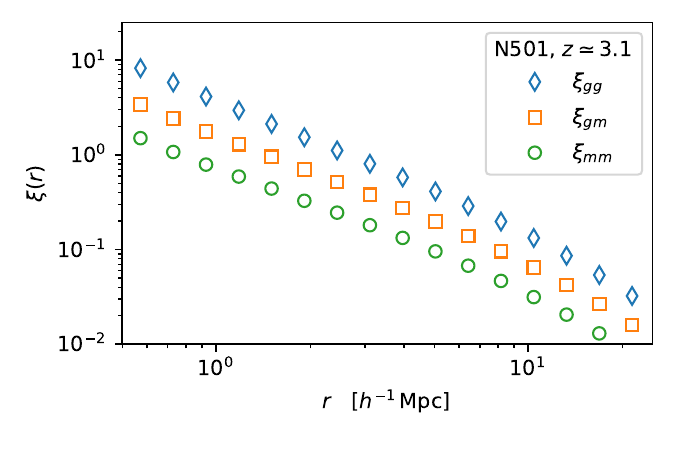}}
    \caption{(Left) The real-space auto and cross power spectra of mock galaxies (with shot-noise suppressed; see text) and the dark matter for the N419 ($z\simeq 2.45$) and N501 ($z\simeq 3.1$) samples (upper and lower).  (Right) The real-space auto and cross correlation functions. }
    \label{fig:real2pt}
\end{figure}

\begin{figure}
    \centering
    \resizebox{\columnwidth}{!}{\includegraphics{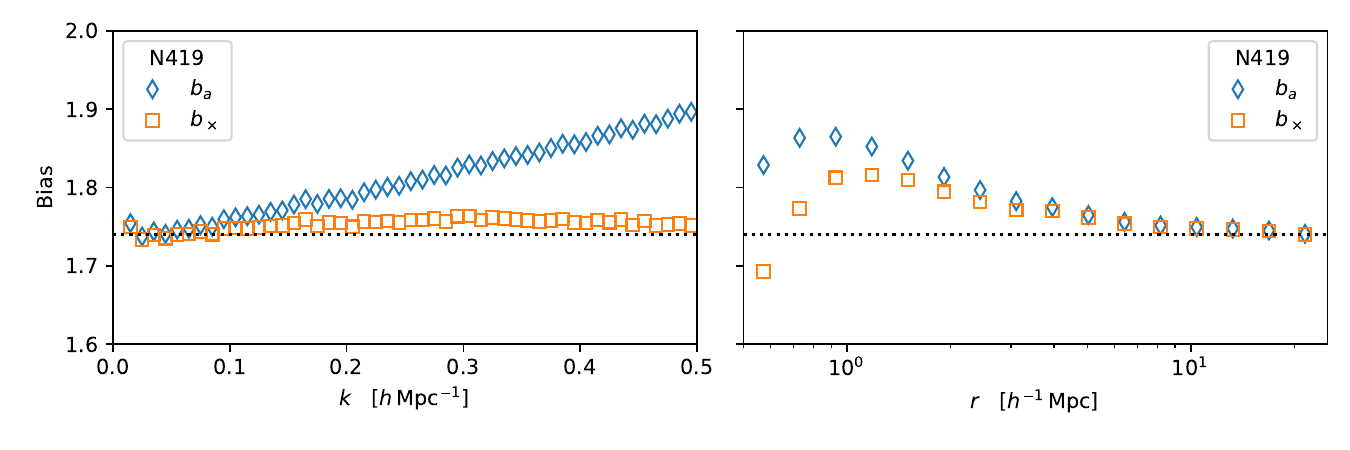}}
    \resizebox{\columnwidth}{!}{\includegraphics{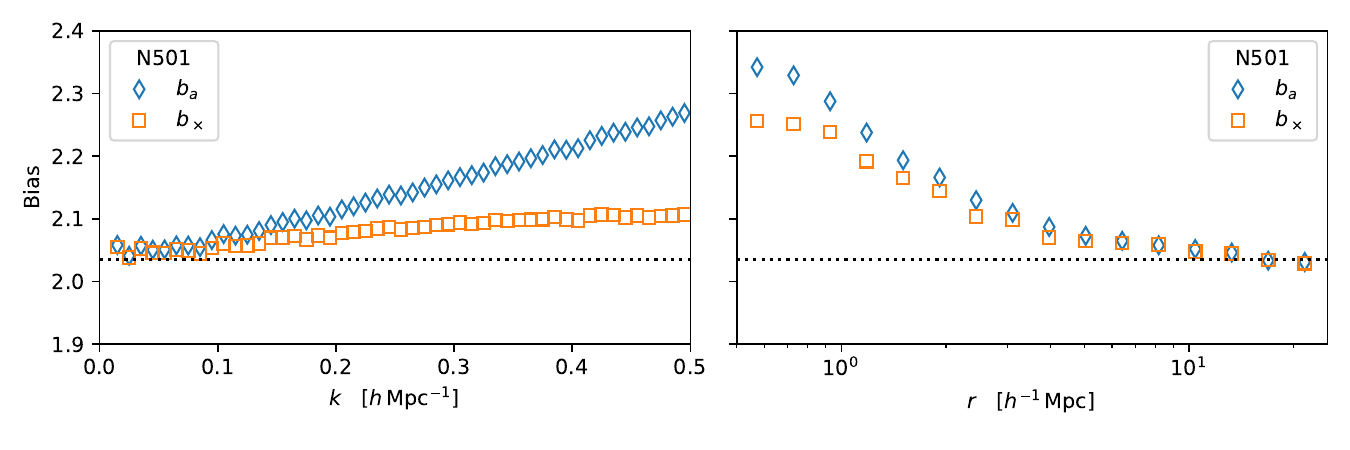}}
    \caption{Scale dependent bias(es) for the N419 ($z\simeq 2.45$) mock sample (upper) and N501 ($z\simeq 3.1$) mock sample (lower).  (Left) the Fourier-space biases, $b_a\equiv\sqrt{P_{gg}/P_{mm}}$ and $b_\times\equiv P_{gm}/P_{mm}$.  (Right) The configuration-space biases, $b_a\equiv\sqrt{\xi_{gg}/\xi_{mm}}$ and $b_\times\equiv \xi_{gm}/\xi_{mm}$.  The non-linear matter 2-point function is used in all cases.  The biases are noticeably scale-dependent on small scales, and $b_a(k)\ge b_\times(k)$ which indicates that $r_{gm}(k)<1$ on small scales. }
    \label{fig:bias}
\end{figure}

Fig.~\ref{fig:real2pt} shows the real-space auto-spectrum of our mock LAEs, as well as the cross-spectrum with the (non-linear) dark matter field, in both Fourier (left) and configuration (right) space.  In Fourier space, to better show the clustered component, we have suppressed shot-noise in the auto-spectrum by not randomly downsampling the mock LAEs in the calculation.  The left panel of Fig.~\ref{fig:bias} shows the implied biases, $b_a=\sqrt{P_{gg}/P_{mm}}$ and $b_\times=P_{gm}/P_{mm}$, as a function of $k$.  At large scales the two estimates agree and are scale-independent with a value consistent with that in our observations.  At small scales (larger $k$) the two display scale dependence and begin to differ as the galaxy field decorrelates from the non-linear matter field ($r=P_{gm}/\sqrt{P_{gg}P_{mm}}=b_\times/b_a<1$).  The galaxy field decorrelates even more rapidly with the linear matter field, or with the initial conditions, and the presence of shot-noise leads to additional decorrelation over that shown in Fig.~\ref{fig:bias}.  The right panel of Fig.~\ref{fig:bias} shows the scale-dependent bias inferred from the configuration space statistics, where we see a similar level of scale-dependent bias but better agreement between $b_a$ and $b_\times$ (though the interpretation of the ratio $b_\times/b_a$ as a correlation coefficient is not valid in configuration space).  Both the scale-dependence of the bias and the decorrelation of the galaxy and matter fields is less pronounced than for a more biased population, such as we might expect of luminous LBGs (e.g.\ Figs~8 and 9 of ref.~\cite{Wilson19}).  This makes LAEs a good candidate for probing primordial physics \cite{Sailer21,Ebina24}, though the samples most useful for cosmology would likely be selected using broader filters than the ODIN filters in order to increase the line-of-sight depth and hence survey volume.

Given the highly complex spectroscopic selection we have not attempted to measure the small-scale, line-of-sight clustering to provide constraints on the virial velocities in the sample (also known as ``fingers of god'' or ``stochastic velocities'').  We anticipate that these will be small, because the large-scale bias implies the halos hosting our galaxies lie primarily on the steeply falling, high-mass tail of the halo mass function.  If satellite LAEs live in even more massive halos than central LAEs their contribution would be highly suppressed (indeed in our best-fitting HOD models the satellite fraction is only a few percent).  If LAE activity is underrepresented in the satellite population this would provide further suppression of the satellite fraction.  To the extent that central galaxies move with their host halo, most of the impact of virial velocities comes from central-satellite pairs and these pairs would be correspondingly suppressed.  This bodes well for using LAEs as cosmological tracers, since virial velocities imply an unavoidable loss of cosmological constraining power on small scales.  It will be very valuable to investigate this issue further with more spectroscopic data.

\begin{figure}
    \centering
    \resizebox{\columnwidth}{!}{\includegraphics{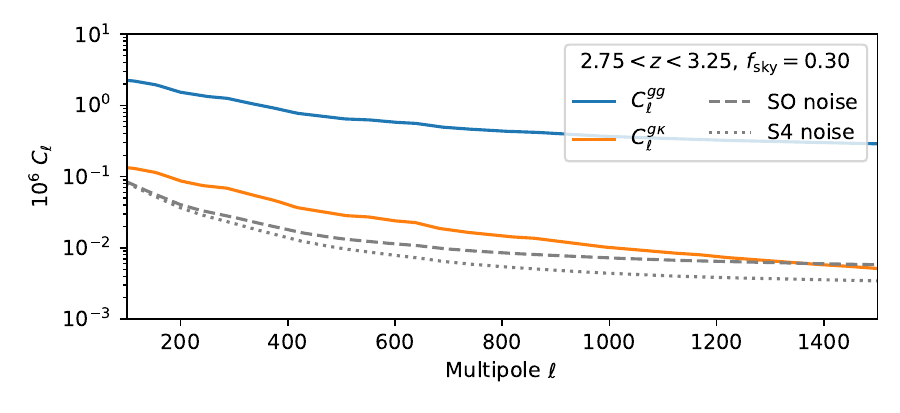}}
    \caption{The projected auto spectrum ($C_\ell^{gg}$) and galaxy-kappa cross spectrum ($C_\ell^{g\kappa}$) assuming LAEs in a slice $2.75<z<3.25$ clustering in the same way as our best-fit LAE sample at $z=3$.  The grey dashed and dotted lines show the error-per-$\ell$-mode on $C_\ell^{g\kappa}$ for noise levels appropriate to Simons Observatory (SO) and CMB-S4 assuming the surveys overlap on 30\% of the sky.  This is to be compared to the signal plotted as the orange line.  Such a cross-correlation would be detected at very high significance by either experiment. }
    \label{fig:CMBlens}
\end{figure}

\subsection{Cross-correlation with CMB lensing}

These LAEs trace large-scale structure in a regime where the kernel of the CMB lensing is still large.  This allows us to measure the cross-correlation between the galaxy and (projected) matter density, which helps to isolate the lensing contributions to a narrow redshift slice. Comparison of the amplitude of clustering inferred by the relativistic tracers (photons) with that inferred from the non-relativistic tracers (LAEs) provides constraints on the theory of gravity.
The real-space power spectra allow us to predict the angular power spectra that would be measured for such galaxies.  For example, for a thin shell of galaxies at distance $\chi_0$ of width $\Delta\chi\ll \chi_0$ we have
\begin{equation}
    C_\ell^{gg} \approx \mathcal{V}^{-1}P_{gg}(k=\ell/\chi_0) \quad , \quad
    C_\ell^{g\kappa} \approx  W^\kappa(\chi_0)\chi_0^{-2}P_{gm}(k=\ell/\chi_0) 
\end{equation}
where $\mathcal{V}=\chi_0^2\,\Delta\chi$ is the volume per steradian and
\begin{equation}
    W^{\kappa}(\chi) = \frac{3}{2} \Omega_{m0}H_0^2 (1+z) \frac{\chi(\chi_{\rm *}-\chi)}{\chi_{\rm *}} 
\end{equation}
is the CMB lensing kernel with $\chi_{\rm *} = \chi(z_{\rm *}{\approx}1100) \approx 9400$ $h^{-1}$Mpc the distance to the surface of last scattering (we have neglected the magnification terms for simplicity).  Note the galaxy auto-spectrum increases as we decrease the width of the shell (and there is less ``washing out'' of the clustering signal) while the cross-spectrum amplitude is independent of $\Delta\chi$.  A very similar expression holds for the cross-spectrum with cosmic shear, i.e.\ galaxy-galaxy lensing, though we don't anticipate having competitive shear measurements at high redshift.

The error on $C_\ell^{g\kappa}$ depends upon $C_\ell^{gg}$, $C_\ell^{g\kappa}$ and $C_\ell^{\kappa\kappa}$ and thus indirectly upon $\Delta\chi$.  A very narrow slice in redshift, such as the ODIN samples analyzed above, has little cross-correlation with CMB lensing since it represents such a tiny fraction of the path traversed by the photons.  However if similar galaxies could be selected over a broader redshift range, e.g.\ by using a series of narrow- or medium-band filters, then very high significance detections could be obtained.  Figure \ref{fig:CMBlens} shows $C_\ell^{g\kappa}$ and its error (per $\ell$-mode) for the best-fitting model at $z=3$ and CMB noise levels appropriate to the Simons Observatory \cite{Lee19} or CMB-S4 \cite{Abazajian19} assuming they overlap on 30\% of the sky and the galaxies cover $\Delta z=0.5$.  In either scenario the cross-spectrum would be detected at very high significance.  For smaller fractions of the sky surveyed the errors scale as $f_{\rm sky}^{-1/2}$, so even smaller surveys would provide highly significant detections.

\subsection{Redshift-space clustering}

\begin{figure}
    \centering
    \resizebox{0.45\columnwidth}{!}{\includegraphics{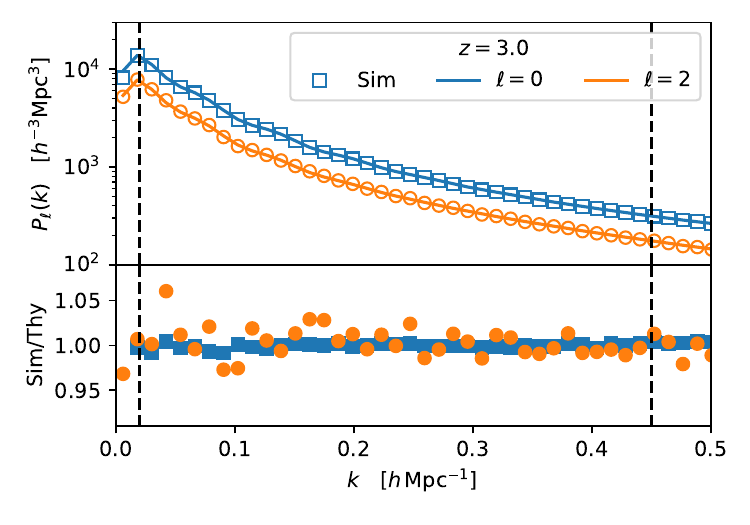}}
    \resizebox{0.45\columnwidth}{!}{\includegraphics{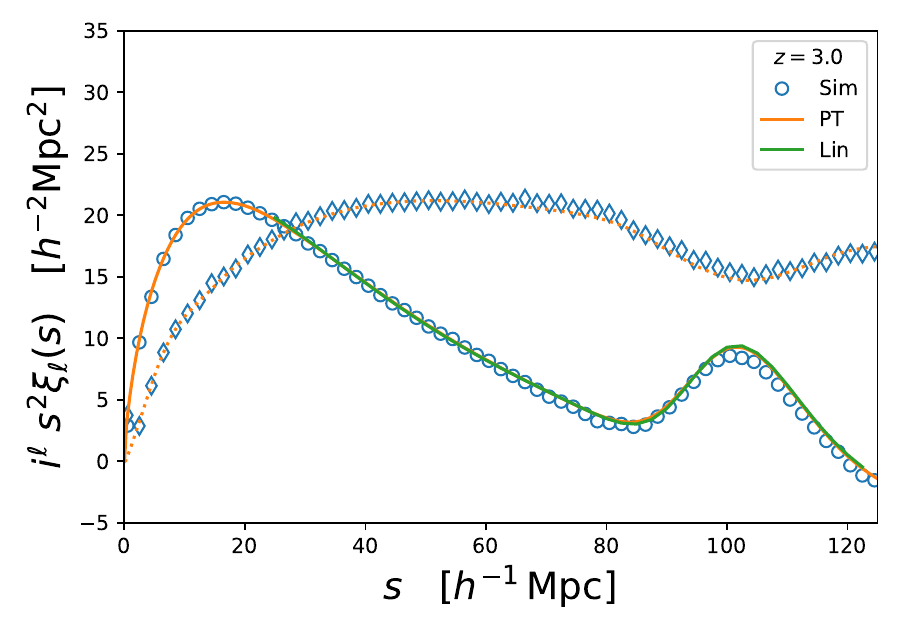}}
    \caption{(Left) The monopole and quadrupole moments of the redshift-space power spectrum measured in the simulation (points; using control variates to reduce noise at large scales), compared to the best-fitting perturbation theory model (lines).  The lower panel shows the ratio of simulation to theory while the vertical dashed lines denote the range of scales fit.  (Right) The monopole and quadrupole of the correlation function measured in the simulation (points) compared to the theoretical prediction from the model of the left panel and linear theory with scale-independent bias.  The broadening of the BAO peak by non-linear evolution is very weak at $z=3$. }
    \label{fig:bao}
\end{figure}

Moving into 3D and redshift space, Fig.~\ref{fig:bao} (left) shows the predicted monopole and quadrupole moments of the redshift-space power spectrum.  The ratio of the two on large scales allows us to measure the growth rate, $f\sigma_8$, with the SNR in this regime being larger the lower the bias and the larger the volume.  The comparison of the two moments at larger $k$ allows constraints on the satellite fraction, virial motions within the halo and scale-dependent bias \cite{Asgari23,Ravi24} though with little constraining power on $f\sigma_8$ \cite{Chen22}.  As discussed previously, since such virial motions represent a key limit to our modeling it would be good to revisit the observational constraints with more data.  In the meantime, Fig.~\ref{fig:bao} also shows a perturbative model \cite{Chen20,Chen21,Maus24} fit to the N-body data over the range $0.02<k<0.45\,h\,\mathrm{Mpc}^{-1}$.  The fit is excellent, and shows that the shift of the non-linear scale to high $k$ at high $z$ means that perturbative models can fit the N-body data over a broad range of scales, providing further support for the forecasting framework used in ref.~\cite{Ebina24}.  The right hand panel of Fig.~\ref{fig:bao} shows the redshift-space clustering in configuration space, with the points being the N-body results and the lines showing linear theory and 1-loop perturbation theory.  The BAO peak visible at $s\sim 100\,h^{-1}\mathrm{Mpc}$ in $\xi_0$ is only modestly broadened by non-linear evolution as the non-linear scale is smaller than the Silk damping scale.

The large-scale bias measurements above also help determine what the simulation requirements would be in order to support a mock catalog effort for future LAE surveys.  Since the central halo occupancy of LAEs is $<1$ even in massive halos, the number density alone cannot be used to infer the characteristic mass of LAE-hosting halos.  That information must come from the large-scale clustering.  Our results (Fig.~\ref{fig:wR_HOD}), and the HOD models that fit them, suggest that the majority of the LAEs we have selected from the ODIN survey reside in halos with $M_h\ge 10^{10.5}\,h^{-1}M_\odot$, though with a tail to lower masses.  This in turn provides a target for the resolution of N-body simulations aimed at modeling this population.  We note that the clustering measured here is lower than the predictions of ref.~\cite{Ravi24}, suggesting that LAEs in those simulations live in more massive halos than LAEs selected from the ODIN survey.

One potential cause for concern in the use of LAEs as tracers of the matter field is the impact of radiative transfer (RT) on the population of galaxies that are selected.  Early work \cite{Zheng11} suggests that RT effects can modulate the number of galaxies in the large-scale structure catalog in a manner that depends upon the local density and line-of-sight velocity divergence.  This latter effect is degenerate with the signal used in redshift-space distortion studies to measure $f\sigma_8$ (as the velocities are `protected' by the equivalence principle and thus independent of galaxy bias).  Follow-up work \cite{Behrens18} argues that such a large effect was due to the limited resolution of those early simulations.  In higher resolution simulations the regions where Ly$\alpha$ photons are produced tend to have higher density, such that the photons escaping from the galaxy have diffused further from the resonance and are less susceptible to the surrounding environment.  Direct observations of these RT effects are currently not available, though the impact of H{\sc i} absorption on LAE clustering has been detected \cite{Momose21}.  Further theoretical and observational work on this is urgently needed, since the impact on forecasts for the science reach of future surveys employing LAEs as tracers is large \cite{Ebina24}.

\section{Conclusions}
\label{sec:conclusions}

Narrow or medium band\footnote{The Rubin Observatory's Legacy Survey of Space and Time (LSST) will reach faint-enough magnitudes to detect a fraction of the LAE population as Lyman break galaxies (LBGs), but optimizing the selection of faint broad-band selected sources which exhibit Ly$\alpha$ emission remains to be done.} surveys provide an efficient route to selecting star-forming galaxies at high redshift \cite{Ouchi20}.  Such galaxies can be valuable tracers of large-scale structure and enable constraints on cosmology and fundamental physics \cite{Ebina24}.  In this paper we have measured the clustering of galaxies selected from the ODIN survey, using follow-up spectroscopy from DESI to refine our selection, to infer the rate of interlopers or contaminants and calibrate the redshift distribution (\S\ref{sec:data}).

Our galaxies reside in thin shells of redshift at $z\simeq 2.45$ and $3.1$.  We measure the angular correlation function of the sample (\S\ref{sec:clustering}), converting from angles to transverse, comoving distance using the mean distance to the shell.  This allows us to side-step the complex spectroscopic selection while paying little penalty in signal-to-noise ratio -- for such shells the clustering on the large scales of most interest for cosmology is dominated by the transverse modes.  We test our pipeline, measure the covariance of our clustering measurement and perform inference using a suite of high-resolution N-body simulations (\S\ref{sec:mocks}) that match the footprint, number density, redshift distribution and clustering of the observed sample.

We find that ODIN-selected LAEs have low clustering amplitudes, indicative of galaxies living in low-mass halos.  Assuming a power-law correlation function with slope $-1.8$, we infer $r_0=(3.0\pm 0.2)h^{-1}$Mpc at both $z\simeq 2.45$ and $3.1$.  In the ``Planck'' \cite{Planck18-VIII} cosmology used in the N-body simulations these correspond to large-scale biases of $1.7\pm 0.2$ and $2.0\pm 0.2$.  We compare our clustering measurements to earlier values from the literature (Fig.\ \ref{fig:lae_r0}), with our results being in the same range as -- though slightly lower than -- previous measurements at comparable redshifts even though our selection differs slightly.

LAEs similar to those studied here make appealing targets for future spectroscopic surveys \cite{Ebina24}.  We use a combination of analytic theory and mock catalogs, constrained by the ODIN-DESI data, to look at the cosmological constraining power of samples with similar properties to our data (\S\ref{sec:forecasts}).  We anticipate that LAEs similar to those studied here could potentially form excellent targets for future, high-$z$, spectroscopic surveys aimed at constraining cosmology and fundamental physics if they could be efficiently selected over a wider redshift range and the impacts of radiative transfer on the selection function were small.

\section{Data availability}
\label{sec:data_availability}

Data from the plots in this paper are available as part of DESI’s Data Management Plan.  The grid of models, predictions and scripts for the analysis and making of the plots can be found at \url{https://github.com/martinjameswhite/AnalyzeLAE} and the remaining data are available at 10.5281/zenodo.11043784.

\acknowledgments
M.W.~is supported by the DOE.
AD's research activities are supported by the NSF NOIRLab, which is managed by the Association of Universities for Research in Astronomy (AURA) under a cooperative agreement with the National Science Foundation. AD's research is also supported in part by Fellowships from the John Simon Guggenheim Memorial Foundation.
EG and KSL acknowledge financial support from the National Science Foundation under Grant Nos. AST-2206222 and 2206705.  EG acknowledges support from the U.S. Department of Energy, Office of Science, Office of High Energy Physics Cosmic Frontier Research program under award No. DE- SC0010008.
ADM was supported by the U.S. Department of Energy, Office of Science, Office of High Energy Physics, under Award Number DE-SC0019022.
This research has made use of NASA's Astrophysics Data System and the arXiv preprint server.
This research is supported by the Director, Office of Science, Office of High Energy Physics of the U.S. Department of Energy under Contract No. DE-AC02-05CH11231, and by the National Energy Research Scientific Computing Center, a DOE Office of Science User Facility under the same contract.


This material is based upon work supported by the U.S. Department of Energy (DOE), Office of Science, Office of High-Energy Physics, under Contract No. DE–AC02–05CH11231, and by the National Energy Research Scientific Computing Center, a DOE Office of Science User Facility under the same contract. Additional support for DESI was provided by the U.S. National Science Foundation (NSF), Division of Astronomical Sciences under Contract No. AST-0950945 to the NSF’s National Optical-Infrared Astronomy Research Laboratory; the Science and Technology Facilities Council of the United Kingdom; the Gordon and Betty Moore Foundation; the Heising-Simons Foundation; the French Alternative Energies and Atomic Energy Commission (CEA); the National Council of Humanities, Science and Technology of Mexico (CONAHCYT); the Ministry of Science and Innovation of Spain (MICINN), and by the DESI Member Institutions: \url{https://www.desi.lbl.gov/collaborating-institutions}. Any opinions, findings, and conclusions or recommendations expressed in this material are those of the author(s) and do not necessarily reflect the views of the U. S. National Science Foundation, the U. S. Department of Energy, or any of the listed funding agencies.

The authors are honored to be permitted to conduct scientific research on Iolkam Du’ag (Kitt Peak), a mountain with particular significance to the Tohono O’odham Nation.


This project is based in part on observations using the Dark Energy Camera (DECam) at the Victor M. Blanco telescope at the NSF Cerro Tololo Inter-American Observatory, NSF NOIRLab (NOIRLab Prop. ID 2020B-0201; PI: K. S. Lee), which is managed by the Association of Universities for Research in Astronomy (AURA) under a cooperative agreement with the U.S. National Science Foundation. DECam was constructed by the Dark Energy Survey (DES) collaboration. Funding for the DES Projects has been provided by the US Department of Energy, the U.S. National Science Foundation, the Ministry of Science and Education of Spain, the Science and Technology Facilities Council of the United Kingdom, the Higher Education Funding Council for England, the National Center for Supercomputing Applications at the University of Illinois at Urbana-Champaign, the Kavli Institute for Cosmological Physics at the University of Chicago, Center for Cosmology and Astro-Particle Physics at the Ohio State University, the Mitchell Institute for Fundamental Physics and Astronomy at Texas A\&M University, Financiadora de Estudos e Projetos, Fundação Carlos Chagas Filho de Amparo à Pesquisa do Estado do Rio de Janeiro, Conselho Nacional de Desenvolvimento Científico e Tecnológico and the Ministério da Ciência, Tecnologia e Inovação, the Deutsche Forschungsgemeinschaft and the Collaborating Institutions in the Dark Energy Survey.

The Collaborating Institutions are Argonne National Laboratory, the University of California at Santa Cruz, the University of Cambridge, Centro de Investigaciones Enérgeticas, Medioambientales y Tecnológicas–Madrid, the University of Chicago, University College London, the DES-Brazil Consortium, the University of Edinburgh, the Eidgenössische Technische Hochschule (ETH) Zürich, Fermi National Accelerator Laboratory, the University of Illinois at Urbana-Champaign, the Institut de Ciències de l’Espai (IEEC/CSIC), the Institut de Física d’Altes Energies, Lawrence Berkeley National Laboratory, the Ludwig-Maximilians Universität München and the associated Excellence Cluster Universe, the University of Michigan, NSF NOIRLab, the University of Nottingham, the Ohio State University, the OzDES Membership Consortium, the University of Pennsylvania, the University of Portsmouth, SLAC National Accelerator Laboratory, Stanford University, the University of Sussex, and Texas A\&M University.

\noindent\texttt{Facilities: Blanco (DECam), Mayall (DESI)}

\appendix

\section{Power-law model}
\label{app:power-law}

It is common in the literature to fit power-law models to LAE clustering data.  To make contact with that practice, here we develop the expressions in the main text for the power-law case.  Specifically we assume $P(k)\propto k^n$, with $-3<n<-1$, and $\Omega_m(z)\approx 1$, appropriate for high redshift observations.  We need to further make an assumption about bias and redshift-space distortions.  We shall take the simplest case with scale-independent bias and linear redshift-space distortions though the validity of these assumptions on small scales is questionable.  With these approximations we can derive expressions for all of our observables analytically.

In terms of the variance per $\log k$, $\Delta_{\rm real}^2(k)=k^3P_{\rm real}(k)/(2\pi^2)\equiv (k/k_\star)^{3+n}$, the multipoles of the redshift-space correlation function are related to the power spectrum multipoles via a Hankel transform
\begin{equation}
    \xi_\ell(r) = i^\ell \int \frac{dk}{k} \Delta_\ell^2(k)\, j_\ell(kr)
    = k_{\ell}\mathcal{I}_{\ell}(n)\,(k_\star r)^{-(3+n)}
\end{equation}
with $j_\ell$ is the spherical Bessel function of order $\ell$ and the $k_\ell$
\begin{equation}
    k_{\rm real} = b^2
    \quad , \quad
    k_0 = \left(b^2+\frac{2}{3}bf+\frac{f^2}{5}\right)
    \quad , \quad
    k_2 = \left( \frac{4}{3}bf + \frac{4}{7}f^2\right)
    \quad , \quad
    k_4 = \frac{8f^2}{35}
\end{equation}
with $f\simeq \Omega_m^{0.55}(z)\approx 1$ and we have defined
\begin{align}
    \mathcal{I}_{\ell}(n) &\equiv i^\ell \int_0^\infty x^{2+n}\, j_\ell(x)\,dx
    \qquad\mathrm{for}\quad
    -3<n<-1 \\
    &= i^\ell\, 2^{1+n}\, \sqrt{\pi} \ \frac{\Gamma([\ell+n+3]/2)}{\Gamma([\ell-n]/2)} .
\end{align}

\begin{figure}
    \centering
    \resizebox{\columnwidth}{!}{\includegraphics{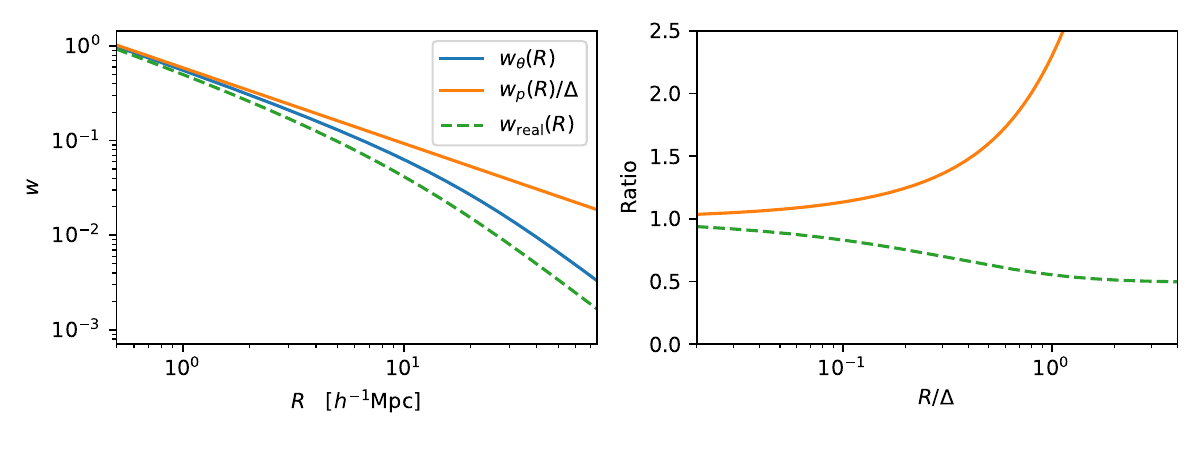}}
    \caption{Power-law example with $\xi_{\rm real}=(1\,h^{-1}\mathrm{Mpc}/r)^{1.8}$, $b=2$ and $\Delta=25\,h^{-1}\mathrm{Mpc}$.  (Left) The full calculation (Eq.~\ref{eqn:wtheta_tophat_shell}; blue) compared to two approximations. The solid orange line shows the approximation $w_\theta\approx w_p/\Delta$ that holds on small scales (Eq.~\ref{eqn:power_law_R_small}) while the dashed green line shows $w_\theta$ computed using the real-space correlation function rather than the redshift-space correlation function.  (Right) The ratio of the two approximations to the full calculation as a function of $R/\Delta$.  Note that both approximations agree with the full result in the limit $R\ll \Delta$, i.e.\ a ``thick'' shell.  Neglecting RSD leads to an underestimate of the clustering on scales approaching the shell thickness while the $w_p$ approximation overestimates power in the same limit.  For lower bias the impact of RSD is larger.}
    \label{fig:powerlaw}
\end{figure}

If we choose $n=-1.2$ then $\xi\propto r^{-1.8}$, which is a frequently used slope as it closely matches the observed correlation functions of several galaxy samples.  We have $\mathcal{I}_0(-1.2)\simeq 1.10725$, $\mathcal{I}_2(-1.2)\simeq -1.66087$ and $\mathcal{I}_4(-1.2)\simeq 1.97229$.  Thus
\begin{align}
    \xi_{\rm real}(r) &= b^2\left(\frac{r_\star}{r}\right)^{1.8}
    \quad\Rightarrow\quad
    \frac{w_p(R)}{R} \simeq 3.679\, b^2 \left( \frac{r_\star}{R} \right)^{1.8}
    \\
    \xi_0(s) &= \left(b^2+\frac{2}{3}bf+\frac{f^2}{5}\right) \left( \frac{r_\star}{s} \right)^{1.8}
    \\
    \xi_2(s) &= -\frac{3}{2}\left( \frac{4}{3}bf + \frac{4}{7}f^2\right) \left( \frac{r_\star}{s} \right)^{1.8}
    \\
    \xi_4(s) &= \frac{57}{32}\, \frac{8f^2}{35} \left( \frac{r_\star}{s} \right)^{1.8}
\end{align}
where $r_\star\simeq (0.945\, k_\star)^{-1}$ is the matter correlation length and we have used $s$ for the redshift-space separation vector to distinguish it from the real-space correlation.

For a top-hat redshift distribution of full-width $\Delta$ the angular correlation function is then
\begin{equation}
  w_\theta(R) = 2\int_0^1 dy\ (1-y)\,
  \sum_{\ell} \xi_\ell(\sqrt{R^2+y^2\Delta^2})\mathcal{L}_\ell\left(\frac{y\Delta}{\sqrt{R^2+y^2\Delta^2}}\right)
\label{eqn:wtheta_tophat_shell}
\end{equation}
where $\mathcal{L}_\ell$ is the Legendre polynomial of order $\ell$.  The integral can be written in closed form using Gaussian hypergeometric functions though the expressions are cumbersome.  If we expand for large $R$ we find
\begin{align}
    w_\theta(R)
    &\to k_0\left(\frac{r_\star}{R}\right)^{1.8}
    \left(1-0.15\frac{\Delta^2}{R^2}+\cdots\right) \nonumber \\
    &+ \frac{3\,k_2}{4}\left(\frac{r_\star}{R}\right)^{1.8}
    \left(1-0.65\frac{\Delta^2}{R^2}+\cdots\right) \nonumber \\
    &+ \frac{171\,k_4}{256}\left(\frac{r_\star}{R}\right)^{1.8}
    \left(1-1.82\frac{\Delta^2}{R^2}+\cdots\right) + \cdots
    \quad\mathrm{as}\ R\to\infty
\end{align}
Note that the contribution from the quadrupole and hexadecapole are similar to that of the monopole, depending upon the value of the bias (and hence the relative sizes of $k_0$ and $k_2$, $k_4$).  For example, for $b=2$ we have $(k_0+3k_2/4+171k_4/256)\simeq 2\,k_{\rm real}$.  Expressions in the limit $\Delta\gg R$ can also be derived simply
\begin{equation}
    w_\theta(R) \to \frac{3.679\, R}{\Delta} \left(\frac{r_\star}{R}\right)^{1.8} \left[
    k_0 - \frac{k_2}{2} + \frac{3 k_4}{8} \right]  
    = \frac{w_p(R)}{\Delta}
    \quad \mathrm{as}\ R\to 0
\label{eqn:power_law_R_small}
\end{equation}
On scales much smaller than the shell width, $\Delta$, the effect of RSD is reduced and $w_\theta(R)\propto w_p(R)$.
We caution that our simple linear model for RSD is likely not appropriate on small scales, and a more reasonable model would include scale-dependent bias, non-linear RSD and the finger-of-god effect (i.e.\ stochastic terms).

To aid with visualizing these results, figure \ref{fig:powerlaw} shows these approximations for a power-law model with $\xi_{\rm real}=(1\,h^{-1}\mathrm{Mpc}/r)^{1.8}$ and $b=2$.  The solid blue line shows the full calculation (Eq.~\ref{eqn:wtheta_tophat_shell}), including (linear) redshift-space distortions and the finite shell width for a shell of width $\Delta=25\,h^{-1}\mathrm{Mpc}$.  The solid orange line shows the approximation $w_\theta\approx w_p/\Delta$ that holds on small scales, as derived above (Eq.~\ref{eqn:power_law_R_small}).  The dashed green line shows $w_\theta$ computed using the real-space correlation function rather than the redshift-space correlation function in Eq.~(\ref{eqn:wtheta_tophat_shell}).  Note that both approximations agree with the full result in the limit $R\ll \Delta$, i.e.\ a ``thick'' shell.  Neglecting redshift-space distortions leads to an underestimate of the clustering on scales approaching the shell thickness while the $w_p$ approximation overestimates power in the same limit.  For lower bias the impact of redshift-space distortions is larger, while for higher bias it is reduced.

\bibliographystyle{JHEP}
\bibliography{main} 
\end{document}